\documentclass[aps,reprint,superscriptaddress,nofootinbib]{revtex4-2}

\usepackage[utf8]{inputenc}
\usepackage[T2A,T1]{fontenc}
\usepackage[russian,main=english]{babel}

\usepackage{newtxtext, tempora, newtxmath, microtype}
\usepackage{accents}
\def\transliteration{latin}
\def\transliteration{cyrillic}
\AtBeginDocument{\DeclareFontFamilySubstitution{T2A}{\rmdefault}{Tempora-TLF}}

\usepackage[per-mode=symbol]{siunitx}

\usepackage{amsfonts, mathrsfs, dutchcal, textgreek, stmaryrd}
\usepackage{mathtools}
\usepackage{tensor}
\allowdisplaybreaks

\usepackage[dvipsnames]{xcolor}
\definecolor{bleu}{RGB}{46,48,141}
\usepackage[colorlinks=true,allcolors=bleu]{hyperref}
\usepackage{cleveref}
\usepackage{orcidlink}

\def\SMF{\textsc{smf}}
\def\ADM{\textsc{adm}}
\def\GRAVITES{\textsc{gravites}}
\def\core{\text{core}}
\def\clad{\text{clad.}}
\def\bulk{\text{bulk}}
\def\intf{\text{intf.}}

\newcommand{\ssqrt}[1]{\sqrt{\vphantom{a^2}\smash{#1}}}
\newcommand{\jump}[1]{\llbracket#1\rrbracket}

\let\t\tensor
\def\p{\partial}

\def\i{\mathrm i}
\def\e{\mathrm e}
\def\d{\mathrm d}
\def\dd{\,\mathrm d}
\def\div{\operatorname{div}}

\def\half{\tfrac{1}{2}}
\def\third{\tfrac{1}{3}}

\def\frameE{e}
\def\frameF{f}

\def\ema{a}
\def\emA{A}
\def\emF{F}
\def\emGd{{\mathfrak G}}
\def\emGt{G}
\def\emGauge{\chi}

\def\amplitude{{\mathscr A}}
\def\jones{{J}}
\def\freq{\omega}
\def\freqK{\omega_*}

\def\neff{\bar n}

\def\Coupling{C}
\def\coupling{c}
\def\gyrotropy{\Gamma}

\def\Metric{g}
\def\Del{\nabla\!}
\def\Riemann{{\underaccent{\bar}{R}}}
\def\Killing{\mathcal K}

\def\metric{l}
\def\del{D\!}
\def\riemann{R}
\def\Weyl{{\underaccent{\bar}{W}}}
\def\lapse{\zeta}
\def\shift{\xi}
\def\twist{\Xi}
\def\DD{\hat D}

\def\baseline{\varGamma}
\def\utangent{t}
\def\unormal{n}
\def\pnormal{\nu}
\def\ubinormal{b}
\def\curvature{\kappa}
\def\torsion{\tau}
\def\dFW{\mathscr D}

\newcommand{\opH}[1]{\t{H}{^{(#1)}}}
\newcommand{\OpH}[1]{\t{\boldsymbol H}{^{(#1)}}}
\newcommand{\opG}[1]{\t{G}{^{(#1)}}}
\newcommand{\OpG}[1]{\t{\boldsymbol G}{^{(#1)}}}
\newcommand{\opGcore}[1]{\t*{G}{_\core^{(#1)}}}
\newcommand{\opGclad}[1]{\t*{G}{_\clad^{(#1)}}}
\newcommand{\OpD}[1]{\t{\boldsymbol \varDelta}{^{(#1)}}}
\newcommand{\hOpD}[1]{\t{\mathbf M}{^{(#1)}}}
\newcommand{\opC}[2]{\t*{C}{_{#1}^{#2}}}

\hypersetup{
	pdftitle={Fiber optics in curved space-times},
	pdfsubject={},
	pdfauthor={Thomas B. Mieling and Mario Hudelist}
}

\begin{document}

\title{Fiber optics in curved space-times}
\author{Thomas B. Mieling~\orcidlink{0000-0002-6905-0183}}
\email{thomas.mieling@univie.ac.at}
\altaffiliation{\newline\newline\newline Based on the article published by APS in \textit{Physical Review Research}, DOI:~\href{https://doi.org/10.1103/PhysRevResearch.7.013162}{PhysRevResearch.7.013162}, under the terms of the \href{https://creativecommons.org/licenses/by/4.0/}{CC BY 4.0 license}. Further distribution of this work must maintain attribution to the author(s) and the published article’s title, journal citation, and DOI.}
\affiliation{University of Vienna, Faculty of Physics and Research Network \textsc{turis}, Boltzmanngasse~5, 1090 Vienna, Austria}
\author{Mario Hudelist~\orcidlink{0000-0001-7498-5939}}
\affiliation{University of Vienna, Faculty of Physics, Boltzmanngasse~5, 1090 Vienna, Austria}

\date{\today}
\begin{abstract}
	Single-mode fibers are used in fiber-optic gyroscopes to measure the Sagnac effect and are planned to be used in forthcoming experiments on the gravitationally induced phase shift in single photons.
	However, current theoretical models of such experiments are limited to ray-optics approximations or, if based on wave optics, to a restricted class of fiber alignments.
	To overcome these shortcomings, this paper develops a comprehensive perturbative scheme to solve for electromagnetic modes, i.e., monochromatic solutions to Maxwell’s equations, of arbitrarily bent step-index fibers in general stationary space-times.
	This leads to transport equations for the electromagnetic phase and polarization that include the gravitational redshift, the Sagnac effect, a generalization of Rytov’s law to curved space, a gravitational Faraday effect in the form of shift-induced gyrotropy, as well as inverse spin Hall effects caused by fiber bending, gravitational acceleration, and space-time curvature.
\end{abstract}
\maketitle

\section{Introduction}

Optical fibers are essential to telecommunication, but also to sensing applications and state-of-the-art quantum optics experiments \cite{2009Senior,2003OptFT...9...57L,2022Optic...9.1238H}.
The latter include interferometric measurements of rotational effects in light-propagation, such as fiber-optic gyroscopes \cite{2013OptFT..19..828L} and multi-photon Sagnac interferometers \cite{2019NJPh...21e3010F,2023PhRvR...5b2005C,2024SciA...10O.215S}, as well as forthcoming tests of the gravitational redshift of single photons and entangled photon pairs in Mach–Zehnder interferometers \cite{2017NJPh...19c3028H,2024Polini}.

Current theoretical models of fiber optics in non-inertial systems and, more generally, curved space-times (to describe gravitational effects using the framework of general relativity) are limited to either strongly simplified models of light propagation or highly restrictive assumptions on the fiber geometry.
Indeed, the vast majority of theoretical models of the Sagnac effect and the gravitational redshift are based on geometrical optics \cite{1967RvMP...39..475P,1975JMP....16..341A,2012CQGra..29v4010Z,2022PhRvA.106c1701M}.
While such models allow for arbitrary fiber geometries, their application to commonly used single-mode fibers (\SMF) is questionable because ray optics is ill-suited for their description even in flat space-time \cite[p.~23]{2009Senior}.
This has motivated the development of more rigorous models based on Maxwell’s equations and their generalization to curved space-times, but due to the complexity of these equations they were successfully applied only to special fiber geometries. For example, Ref.~\cite{1982ApOpt..21.1400L} computed the Sagnac effect in planar fiber loops (using a Gaussian approximation of the transverse field profile, thereby avoiding mathematical complications arising from discontinuities in the \SMF’s index profile) and Ref.~\cite{2020CQGra..37v5001M} gave a solution for straight fibers (without Gaussian approximations). Similarly, rigorous analyses of the gravitational redshift in \SMF\ are currently limited to straight fibers at constant gravitational potentials in post-Newtonian metrics \cite{2018CQGra..35x4001B,2022PhRvA.106f3511M}, while experimental proposals suggest using fiber-optic spools instead \cite{2017NJPh...19c3028H}. Likewise, current theoretical models of optical solitons in the Schwarzschild geometry are limited so a small set of fiber alignments \cite{2023CQGra..40n5008S,2024NJPh...26h3010B}.

Recent theory developments allow overcoming these limitations of fiber-optics calculations. Specifically, Ref.~\cite{2023PhRvR...5b3140M} developed a perturbative scheme to determine phase shifts and polarization transport laws for bent \SMF\ in flat space-time (neglecting bending losses), thus providing the first rigorous derivation of Rytov’s law, which had already been verified experimentally \cite{1984OQEle..16..455R,1986PhRvL..57..937T} but had previously been derived only using ray optics methods \cite{1926Bortolotti,1938Rytov_\transliteration,1989Rytov_english} or approximation schemes that neglected the internal structure of \SMF\ \cite{1987Natur.326..277B,2018PhRvA..97c3843L,2019PhRvA.100c3825L}.

The present paper combines and extends the methods of Refs.~\cite{2023PhRvR...5b3140M,2023Mieling,2024Hudelist} to provide a comprehensive model of light-propagation in single-mode step-index fibers in general stationary space-times. The result of the calculation is a series of transport laws, formulated in equations \eqref{eq:summary:transport frequency} to \eqref{eq:summary:transport polarization}, that describe the variation of the frequency, wavelength, phase, and polarization along a bent fiber, respectively.

The notation used in this paper is summarized in \cref{app:notation}.

\section{Stationary Fibers}
\label{sec:setup}

\begin{figure}
	\centering
	\includegraphics[width=0.9\columnwidth]{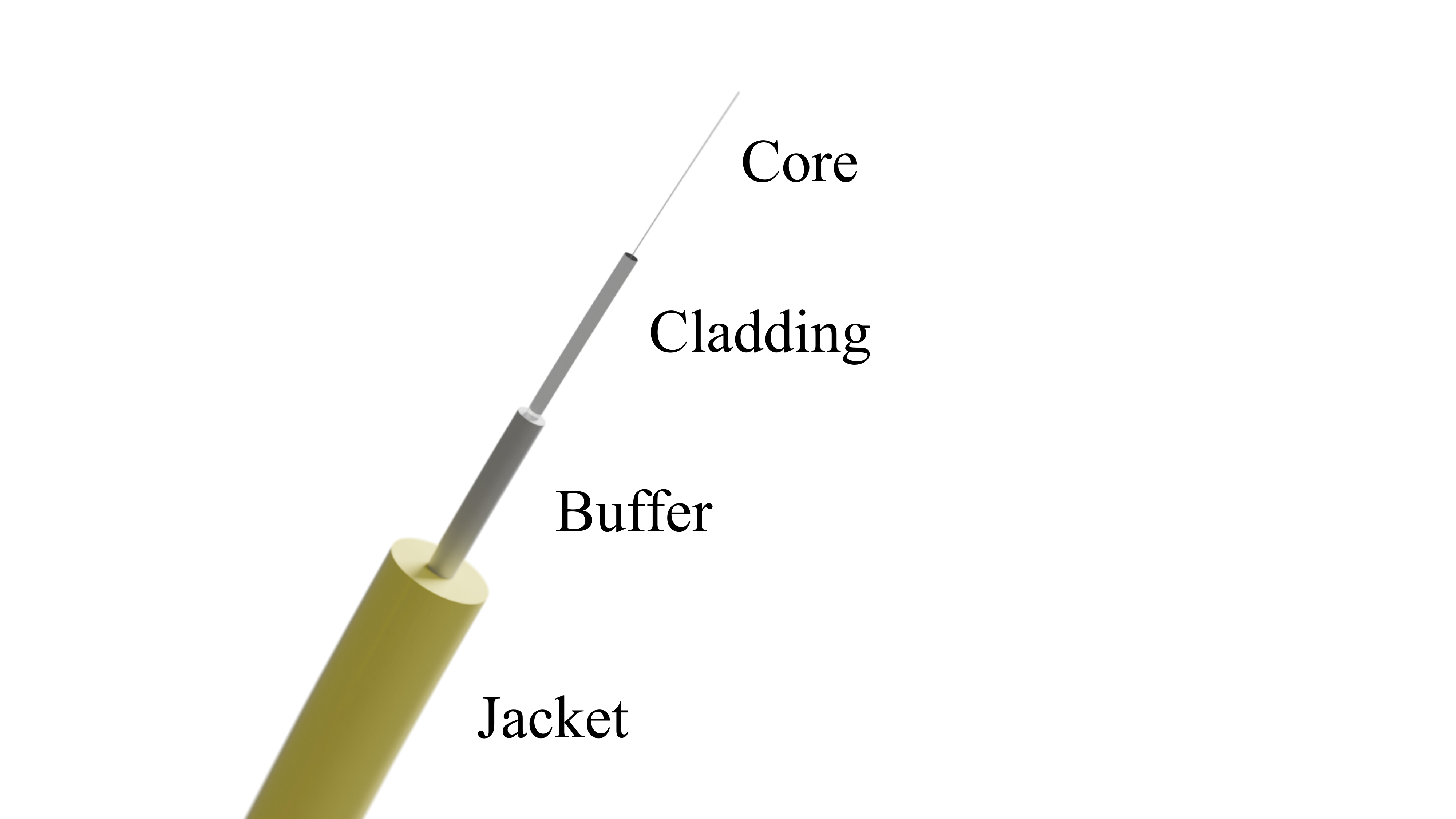}
	\caption{%
		Schematic representation of the internal structure of single-mode step-index fibers.
		The electromagnetic modes are supported in the core (typical diameters $ 2 \varrho < \SI{10}{\micro\meter}$) and decay at exponential rates in the cladding (diameters on the order of \SI{100}{\micro\meter}).
		For the purpose of solving Maxwell’s equations, the protective layers (buffer and jacket with typical outer diameters on the order of \SI{1}{\milli\meter}) are thus irrelevant.
	}
	\label{fig:fiber structure}
\end{figure}

Single mode fibers (\SMF) consist of an optically dense core within an optically less dense cladding.
Whereas nanofibers consist of a glass core with ambient air acting as the cladding, the core and cladding of typical \SMF\ are made of fused silica and are shielded by additional protective layers, see \cref{fig:fiber structure}.
This section models such fibers in stationary curved space-times, thus providing a basis for the description of light propagation within such fibers.

\begin{figure}
	\centering
	\includegraphics[width=0.9\columnwidth]{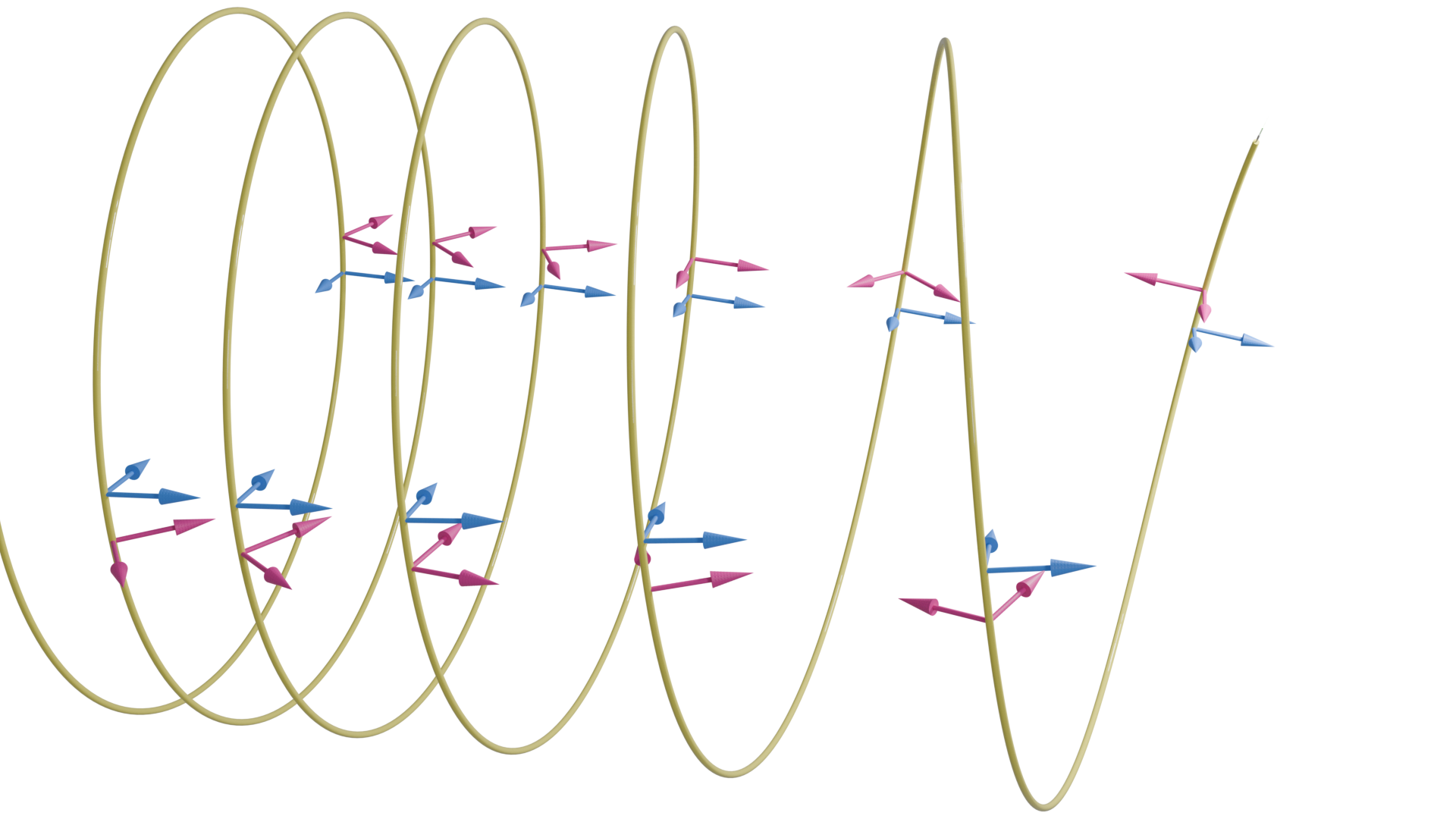}
	\caption{%
		Schematic representation of a coiled optical fiber together with the Frenet–Serret frame (blue) and the Bishop frame (magenta).
		Whereas the Frenet–Serret frame is obtained by differentiating the curve’s unit tangent, the Bishop frame is obtained by Fermi–Walker transport.
		\Cref{eq:Frenet Serret} implies that the angle between the two frames increases with arc length at a rate that is equal to the curve’s torsion.
	}
	\label{fig:fiber geometry}
\end{figure}

Stationary space-times are described by a metric $\t\Metric{_\mu_\nu}$ that admits a timelike Killing vector field $\t\Killing{^\mu}$ \cite[Chap.~18]{2003ExactSolutions}.
The dimensionless norm $\lapse = \sqrt{- \t\Metric{_\mu_\nu} \t\Killing{^\mu} \t\Killing{^\nu}} / c$ will henceforward be referred to as the lapse.
Locally, there exist coordinates $(t, \t x{^i})$ with respect to which $\t\Killing{^\mu}$ takes the form $(\t\Killing{^\mu}) = (1, \boldsymbol 0)$.
The one-form $\t\Killing{_\mu} \equiv \t\Metric{_\mu_\nu} \t\Killing{^\nu}$ can then be parametrized as $(\t\Killing{_\mu}) = c^2 \lapse^2(-1, \t\shift{_i} / c)$, where $\t\shift{_i}$ will be referred to as the shift.%
\footnote{%
	Apart from allowing arbitrary $t$-independent transformations of the spatial coordinates $\t x{^i}$, the setup also allows for a certain “gauge freedom” that arises from the possibility of redefining the temporal coordinate as $t' = t + \theta$ provided that $\theta$ is any smooth function of the spatial coordinates $\t x{^i}$. This induces a “gauge transformation” of the shift of the form $\t*\shift{_i^\prime} = \t\shift{_i} + c \t\p{_i}\! \theta$.%
	\label{ft:gauge transform shift}%
}
Finally, defining the spatial metric $\t\metric{_\mu_\nu}$ relative to $\t\Killing{^\mu}$ as $\t\metric{_\mu_\nu} = \t\Metric{_\mu_\nu} + \t\Killing{_\mu} \t\Killing{_\nu} / (c \lapse)^2$ yields the decomposition \cite[Sect.~9.6]{2001Rindler}
\begin{align}
	\label{eq:metric general}
	\Metric
		&= - \lapse^2 (c \d t - \t\shift{_i} \t{\d x}{^i})^2
		+ \t\metric{_i_j} \t{\d x}{^i} \t{\d x}{^j}\,.
\end{align}
Compared to the \ADM\ metric $\t*g{^\ADM_i_j} = \t\metric{_i_j} - \lapse^2 \t\shift{_i} \t\shift{_j}$, which describes the geometry of the level-sets $t = \text{constant}$ \cite{1959PhRv..116.1322A}, the spatial metric $\t\metric{_i_j}$ describes the geometry of events that are simultaneous relative to the four-velocity $\t u{^\mu} = \t\Killing{^\mu} / \lapse$ \cite[§~84]{Landau2}.
Moreover, the spatial metric $\t\metric{_i_j}$ as defined here is invariant under the gauge transformations described in footnote~\ref{ft:gauge transform shift}, whereas $\t*g{^\ADM_i_j}$ is not invariant under such transformations.
Based on the above decomposition of the space-time metric, the shape of optical fibers “at rest” can be modeled in the spatial geometry that is described by the metric $\t\metric{_i_j}$. To this end, it is beneficial to use spatial coordinates that are adapted to the fiber’s baseline $\baseline$.
Denoting by $\t\utangent{^i} = \d\t\baseline{^i}(s)/\d s$ the unit tangent, where $s$ is the arc length, the principal normal is $\t\pnormal{^i} = \t{D\!}{_s} \t\utangent{^i}$ where $\t{D\!}{_i}$ is the spatial Levi-Civita derivative. In regions of non-zero curvature $\curvature = (\t\metric{^i^j}\t\pnormal{_i} \t\pnormal{_j})^{1/2}$, the unit normal is given by $\t\unormal{^i} = \t\pnormal{^i}/\curvature$, and the unit binormal is $\t\ubinormal{^i} = \t\epsilon{^i_j_k} \t\utangent{^j} \t\unormal{^k}$, where $\t\epsilon{_i_j_k}$ is the spatial Levi-Civita (pseudo-)tensor.
The frame $(\t\utangent{^i}, \t\unormal{^i}, \t\ubinormal{^i})$ satisfies the Frenet–Serret equations \cite[p.~34]{1999Spivak_2}
\begin{align}
	\label{eq:Frenet Serret}
	\t\dFW{_s} \t\utangent{^i} &= 0\,,
	&
	\t\dFW{_s} \t\unormal{^i} &= + \torsion \t\ubinormal{^i}\,,
	&
	\t\dFW{_s} \t\ubinormal{^i} &= - \torsion \t\unormal{^i}\,,
\end{align}
where $\tau = + \t\ubinormal{_i} \t\del{_s} \t\unormal{^i} = - \t\unormal{_i} \t\del{_s} \t\ubinormal{^i}$ is the torsion of $\baseline$, and $\t\dFW{_s}$ denotes the Fermi–Walker derivative
\begin{align}
\label{eq:Fermi Walker}
\begin{split}
	\t\dFW{_s} \t w{^i}
		&= \t\del{_s} \t w{^i}
		+ \t\utangent{^i} \t\pnormal{_k} \t w{^k}
		- \t\pnormal{^i} \t\utangent{_k} \t w{^k}
		\\&
		\equiv \t\del{_s} \t w{^i}
		- \curvature \t\epsilon{^i_j_k} \t\ubinormal{^j} \t w{^k}
		\,.
\end{split}
\end{align}
References~\cite{1987Natur.326..277B,2018PhRvA..97c3843L,2019PhRvA.100c3825L} described light propagation in optical fibers using the Frenet–Serret basis. However, since this frame is undefined at points of vanishing curvature the following analysis uses, instead, an orthonormal frame $(\t\utangent{^i}, \t*\frameF{_1^i}, \t*\frameF{_2^i})$ satisfying Bishop’s equations \cite{1975Bishop}
\begin{align}
	\label{eq:frame Bishop transport}
	\t{\mathscr D}{_s} \t*\frameF{_1^i} &= 0\,,
	&
	\t{\mathscr D}{_s} \t*\frameF{_2^i} &= 0\,.
\end{align}
\Cref{fig:fiber geometry} provides a graphical illustration of the transverse Frenet–Serret frame $(\t\unormal{^i}, \t\ubinormal{^i})$ and the Fermi–Walker-transported Bishop frame $(\t*\frameF{_1^i}, \t*\frameF{_2^i})$ for a coiled fiber.
The latter frame can be used to erect spatial Fermi coordinates $(s, \t x{^1}, \t x{^2}) \equiv (s, \t x{^\alpha})$, see, e.g., Ref.~\cite[Appendix A]{2023PhRvR...5b3140M}.
In these coordinates, the physical distance of a point $(s, \t x{^\alpha})$ from the baseline is given by $\ssqrt{\t{\delta\!}{_\alpha_\beta} \, \t x{^\alpha} \t x{^\beta}}$, so circles and cylinders around the baseline are accurately represented.
Moreover, the series-expansion of the spatial metric around the baseline reads
\begin{align}
	\label{eq:spatial metric expansion}
	\begin{split}
		\metric = {}&
			[ (1 - \t\pnormal{_\alpha} \t x{^\alpha})^2 - \t\riemann{_s_\alpha_s_\beta} \t x{^\alpha} \t x{^\beta} ] \d s^2
			\\&+
			\tfrac{4}{3} \t\riemann{_s_\alpha_\beta_\gamma} \t x{^\alpha} \t x{^\beta} \d s \, \d\t x{^\gamma}
			\\&+
			[\t\delta{_\alpha_\beta} - \third \t\riemann{_\alpha_\gamma_\beta_\delta} \t x{^\gamma} \t x{^\delta} ] \d\t x{^\alpha} \, \d\t x{^\beta}
			\\&+
			O[(\t\delta{_\alpha_\beta} \t x{^\alpha} \t x{^\beta})^{3/2}]\,,
	\end{split}
\end{align}
where all components of the spatial Riemann tensor $\t\riemann{_i_j_k_l}$ are evaluated on the baseline $\baseline$.
The deviations from the canonical form of the metric arising here have the same structure as in the Lorentzian case \cite{1963JMP.....4..735M,1978PhRvD..17.1473N}.
Similarly to the spatial metric $\t\metric{_i_j}$, the lapse can be expanded as a Taylor series in the form
\begin{align}
	\label{eq:lapse expansion}
	\begin{split}
		\lapse(s, \t x{^\alpha}) ={}&
			\lapse(s)
			+ \t\lapse{_,_\alpha}(s) \t x{^\alpha}
			\\&
			+ \half \t\lapse{_,_\alpha_\beta}(s) \t x{^\alpha} \t x{^\beta}
			+ O[(\t\delta{_\alpha_\beta} \t x{^\alpha} \t x{^\beta})^{3/2}]\,,
	\end{split}
\end{align}
and analogous equations apply to the components $\t\shift{_i}$ of the shift.%
\footnote{One generally has $\t\lapse{_,_\alpha} \equiv \t\p{_\alpha} \lapse = \t{D\!}{_\alpha} \lapse$, but in the considered coordinates it also holds that $\t\lapse{_,_\alpha_\beta} \equiv \t\p{_\alpha}\t\p{_\beta} \lapse = \t{D\!}{_\alpha} \t{D\!}{_\beta} \lapse$ and $\t\shift{_\alpha_,_\beta} \equiv \t\p{_\beta} \t\shift{_\alpha} = \t{D\!}{_\beta} \t\shift{_\alpha}$. This is because (up to symmetry) the only non-trivial spatial Christoffel symbols along the baseline $\baseline$ are $\t\Gamma{_\alpha_s_s} = + \t\pnormal{_\alpha}$ and $\t\Gamma{_s_s_\alpha} = - \t\pnormal{_\alpha}$.}
These series expansions provide a good approximation to $\t\Metric{_\mu_\nu}$ within the optical fiber whenever the characteristic length scales of the space-time geometry are significantly larger than the fiber’s diameter.
In practice, this is always satisfied.

Due to the accurate representation of cylinders by Fermi coordinates, step-index fibers can be described as in Euclidean space.
In particular, for fibers with translation-invariant and circularly symmetric cross-sections, the fiber core corresponds to a disk in the $(\t x{^1}, \t x{^2})$-plane, and the additional layers (cladding, buffer, and jacket) form concentric annuli.

To describe light propagation in \SMF, one must describe not only the geometry of its internal structure but also its electromagnetic properties.
As is common practice, the following calculations model such fibers as linear, isotropic and non-magnetic dielectrics with constant real refractive indices $\t n{_i}$ in each layer \cite[Sect.~2.4.3]{2009Senior}.
Additionally, due to the rapid radial decay of the electromagnetic field in the cladding, its outer radius is set to infinity in the following calculations \cite[Sect.~3.1]{2005Liu}.
Under these assumptions, the constitutive equation relating the electromagnetic field-strength $\t\emF{_\mu_\nu}$ (a two-form) to the excitation $\t\emGd{^\mu^\nu}$ (a bivector-density) can be written as $\t\emGd{^\mu^\nu} = \half \t{\mathfrak X}{^\mu^\nu^\rho^\sigma} \t\emF{_\rho_\sigma}$ \cite[Sect.~VI.2]{1962Post}, in which the constitutive tensor density $\t{\mathfrak X}{^\mu^\nu^\rho^\sigma}$ can be expressed in terms of Gordon’s contravariant optical metric $\t{\tilde g}{^\mu^\nu} = \t\metric{^\mu^\nu} - (n^2/c^2) \t u{^\mu} \t u{^\nu}$ as \cite{1923AnP...377..421G}
\begin{align}
	\t{\mathfrak X}{^\mu^\nu^\rho^\sigma}
		&= \sqrt{- g} [
			  \t{\tilde g}{^\mu^\rho} \t{\tilde g}{^\nu^\sigma}
			- \t{\tilde g}{^\mu^\sigma} \t{\tilde g}{^\nu^\rho}
		]\,.
\end{align}
In the following, electromagnetic modes are computed in terms of the four-potential $\t\emA{_\mu}$.
Since Maxwell’s equations do not directly lead to well-posed evolution equations for $\t\emA{_\mu}$, the approach taken here is to work, instead, with the field equation of Ref.~\cite{2022PhRvA.106f3511M}, namely
\begin{align}
	\label{eq:field equation extended}
	\t{\nabla\!}{_\mu} \t\emGt{^\mu^\nu} + \t{\tilde g}{^\mu^\nu} \t{\nabla\!}{_\mu} \emGauge = 0\,,
\end{align}
where $\t\emGt{^\mu^\nu} = \t{\tilde g}{^\mu^\rho} \t{\tilde g}{^\nu^\sigma} \t{(\d\emA)}{_\rho_\sigma}$ and $\emGauge = \t\Del{_\mu}(\t{\tilde g}{^\mu^\nu} \t\emA{_\nu})$.
Contrary to Maxwell’s equations, \cref{eq:field equation extended} constitutes a well-posed evolution equation for the potential, and if $\t\emA{_\mu}$ is a solution to \cref{eq:field equation extended} with $\emGauge = 0$, then $\t\emF{_\mu_\nu} = \t{(\d\emA)}{_\mu_\nu}$ and $\t\emGd{^\mu^\nu} = \sqrt{-g} \t\emGt{^\mu^\nu}$ satisfy Maxwell’s equations, together with the constitutive equation, in the form
\begin{align}
	\t{(\d\emF)}{_\mu_\nu_\rho} &= 0\,,
	&
	\t{(\div \emGd)}{^\mu} &= 0\,,
	&
	\t\emGd{^\mu^\nu} &= \half \t{\mathfrak X}{^\mu^\nu^\rho^\sigma} \t\emF{_\rho_\sigma}\,.
\end{align}
This procedure is analogous to the method of solving the “relaxed Einstein equations” described in Ref.~\cite[Sect.~6.2]{2014PoissonWill}.

In the present case, obtaining exact closed-form solutions to \cref{eq:field equation extended} is infeasible.
The following analysis is thus based on a perturbative expansion in a parameter $\varepsilon \ll 1$ that derives from the assumption that certain geometric quantities are small and gradually varying when compared to a typical length scale $\ell_0$ associated to electromagnetic radiation in \SMF\ (such as the fiber’s core radius $\varrho$ or the vacuum wavelength $2 \pi c / \freq$).
Specifically, a dimensionless function $\varphi$ along the fiber’s baseline $\baseline$ will be said to be $k$-gradual if it can be written in the form $\varphi(s) = \varepsilon^k \mathring \varphi(\varepsilon s / \ell_0)$, where $\mathring \varphi$ is a bounded function all of whose derivatives do not exceed the order of magnitude of $\mathring \varphi$ itself.
The following analysis assumes $\lapse$ to be $0$-gradual, that $\ell_0 \t\lapse{_,_i}$, $\t\shift{_i}$, $\ell_0 \t\pnormal{_i}$, are $1$-gradual, and $\ell_0^2 \t\lapse{_,_i_j}$, $\ell_0 \t\shift{_i_,_j}$, $\ell_0^2 \t\riemann{_i_j_k_l}$ are assumed to be $2$-gradual.%
\footnote{The underlying heuristic behind these assumptions is that the fiber’s curvature and the ambient geometry have characteristic length scales $\ell$ that far exceed the optical length scale $\ell_0$. All functions listed above, except for $\t\shift{_i}$ are thus naturally of the form described above with $\varepsilon$ replaced by $\varepsilon' = \ell_0 / \ell \ll 1$.
For technical reasons, the following analysis also requires the (dimensionless) shift $\t\shift{_i}$ to be small. Assuming a uniform bound $|\t\shift{_i}| < \varepsilon'' \ll 1$, these assumptions can be cast into the form given above by setting $\varepsilon = \operatorname{max}\{\varepsilon', \varepsilon''\}$.}
More explicitly, these assumptions require the existence of a parameter $\varepsilon \ll 1$ such that
\begin{subequations}
\label{eq:smallness}
\begin{align}
	\label{eq:smallness:1}
	\varrho \omega / c &= O(1)\,,
	&
	\lapse &= O(1)\,,
	\\
	\label{eq:smallness:2}
	\t\shift{_i} &= O(\varepsilon)\,,
	&
	\varrho \t\pnormal{_i} &= O(\varepsilon)\,,
	\\
	\label{eq:smallness:3}
	\varrho^2 \t\riemann{_i_j_k_l} &= O(\varepsilon^2)\,,
	&
	\varrho \del{_i} f &= f\, O(\varepsilon)\,,
\end{align}
\end{subequations}
where, in the last expression, $f$ ranges over the functions $\lapse, \t\shift{_i}, \t\riemann{_i_j_k_l}$ and their derivatives, all of which are gradually varying in the sense of depending on the arc length $s$ solely through $\varepsilon s / \ell_0$. \Cref{eq:smallness:1} expresses that the optical wavelength is on the order of the fiber core radius and that the lapse function can be of order unity (it is equal to $1$ for inertial coordinates in flat space-time). Further, \cref{eq:smallness:2} requires the rotation rate of the fiber to be small (bounding the shift $\t\shift{_i}$) and the fiber bending radius to be large compared to the core radius (bounding $\varrho \t\pnormal{_i}$ and hence also $\varrho \curvature$).
Finally, \cref{eq:smallness:3} requires the spatial curvatures to be one order smaller than the previously considered quantities (so that $\t\riemann{_i_j_k_l}$ is at most on the order of $\t\pnormal{_i} \t\pnormal{_j}$), and that taking transverse derivatives of the discussed quantities increases the order in $\varepsilon$ by one.
These assumptions generalize the weak-bending assumption formulated in Ref.~\cite{2023PhRvR...5b3140M} and are analogous to the scaling hypotheses underlying models of the gravitational and electromagnetic self-forces \cite{2008CQGra..25t5009G,2009PhRvD..80b4031G}.
While the smallness assumptions in \cref{eq:smallness} are satisfied in typical \SMF, they can be violated in tightly bent nanofibers \cite{2003Natur.426..816T} and would, in principle, also fail for relativistic rotational velocities or strong tidal forces with curvature radii below a micrometer (though such violations do not occur in practical applications).

\section{Perturbative Scheme}
\label{sec:perturbative scheme}

To solve, perturbatively, for the electromagnetic modes in the setup described above, it is useful to introduce dimensionless cylindrical coordinates $(\t x{^0}, \sigma, r, \vartheta)$ via
\begin{subequations}
\begin{align}
	t &= \ell_0 \t x{^0} / c\,,
	&
	\t x{^1} &= \ell_0 r \cos \vartheta\,,
	\\
	s &= \ell_0 \sigma\,,
	&
	\t x{^2} &= \ell_0 r \sin \vartheta\,,
\end{align}
\end{subequations}
where $\ell_0$ is the optical length scale introduced in the previous section.
For definiteness, for \SMF\ with translation-invariant circular cross-sections, maximal simplification of the equations is achieved by setting $\ell_0$ to the core radius $\varrho$ (on the order of micrometers).
Moreover, since the calculation makes use of Fourier series in $\vartheta$, it is advantageous to define the complex frame
\begin{align}
	\label{eq:complex frame}
	\t*\frameF{_\pm^i}
		&= \tfrac{1}{\sqrt 2}(\t*\frameF{_1^i} \mp \i \t*\frameF{_2^i})\,,
\end{align}
with the normalization chosen such that $\t\metric{_i_j} \t*\frameF{_+^i} \t*\frameF{_-^j} = 1$, since the number of $+$ and $-$ subscripts in a term determines its “Fourier index,” i.e., the integer $m$ in the expression $\e^{\i m \vartheta}$. For example, setting $\t\pnormal{_\pm} = \t\pnormal{_i} \t*\frameF{^i_\pm}$, one has $\t\pnormal{_\alpha} \t x{^\alpha} / \ell_0 = (r/\sqrt 2)(\t\pnormal{_+} \e^{+\i\vartheta} + \t\pnormal{_-} \e^{-\i\vartheta})$.

Although it is common practice to study Maxwell’s equations in optical fibers using the coordinate coframe $\t{\d x}{^\mu}$ associated to cylindrical coordinates \cite{2005Liu,2017NJPh...19c3028H}, the approach taken here is to use, instead, the complex coframe
\begin{subequations}
\label{eq:complex coframe}
\begin{align}
	\t\frameE{^0}
		&= \lapse(s) [\t{\d x}{^0} - \t\shift{_\sigma} \d \sigma ]\,,
	&
	\t\frameE{^\sharp}
		&= \tfrac{1}{\sqrt 2}(\d r - \i r \d \vartheta)\,,
	\\
	\t\frameE{^\parallel}
		&= \d \sigma\,,
	&
	\t\frameE{^\flat}
		&= \tfrac{1}{\sqrt 2}(\d r + \i r \d \vartheta)\,,
\end{align}
\end{subequations}
since this leads to fully decoupled field equations at leading order, cf.\ Refs.~\cite{2022PhRvA.106f3511M,2023PhRvR...5b3140M,2023arXiv230212729S}.%
\footnote{%
	The complex frame \eqref{eq:complex frame} and the complex coframe \eqref{eq:complex coframe} are complementary in the sense that $\t*\frameF{_+^i}$ and $\t*\frameF{_-^i}$ are defined along the baseline $\baseline$ ($r = 0$), whereas $\t*\frameE{^\sharp_i}$ and $ \t*\frameE{^\flat_i}$ are defined away from $\baseline$ ($r > 0$).}
Monochromatic modes, i.e., traveling waves with constant Killing frequency $\freqK$, can be computed perturbatively using the ansatz $\t\emA{_\mu} = \t\emA{_b} \t\frameE{^b_\mu}$ with
\begin{align}
	\label{eq:ansatz amplitude phase}
	\t\emA{_b}
		&= \t\ema{_b}(r,\vartheta,\varsigma) \, \e^{\i \psi}\,,
\end{align}
where $\psi$ is the phase function
\begin{align}
	\label{eq:ansatz phase}
	\psi
		&= - \freqK \left[
			t - \t\shift{_\alpha}(s) \t x{^\alpha} / c
			- \int\hspace{-0.25em} \t\shift{_s}(s) \frac{\d s}{c}
		\right]
		\pm \!\int\hspace{-0.25em} \beta(s) \d s\,,
\end{align}
with the local propagation parameter $\beta > 0$ (such that the sign determines the direction of light propagation) and the amplitudes $\t\ema{_b}$ being expanded as
\begin{align}
	\label{eq:ansatz beta}
	\beta
		&= \sum_{j \in \mathbf N} \varepsilon^j \beta^{(j)}(\varsigma)\,,
	\\
	\label{eq:ansatz a}
	\t\ema{_b}
		&= \sum_{j \in \mathbf N} \sum_{m \in \mathbf Z} \varepsilon^j \e^{\i m \theta} \t*\ema{_b^{(j,m)}}(r, \varsigma)\,.
\end{align}
In these and the following equations, $\sigma$ and $\varsigma = \varepsilon \sigma$ are treated as “independent variables” in the sense of the multiple-scales method \cite[Chap.~11]{1978BenderOrszag}.
The explicit form of the phase function $\psi$ can be motivated using geometrical optics: interpreting $\t k{_\mu} =\t{(\d \psi)}{_\mu}$ as the local wave-covector along the baseline, one has $\t k{_\mu} = - (\omega / c^2) \t u{_\mu} \pm \beta \t{(\d s)}{_\mu}$, where $\freq = \freqK / \lapse$ is the frequency relative to the medium.
The quantity $n_\text{ph} = c \beta / \omega$ can thus be interpreted as the mode’s phase index that must be determined from the field equations (it typically lies between the core index $n_1$ and the cladding index $n_2$). In the following, the leading-order contribution $\neff = c \beta^{(0)}/\omega$ will be referred to as the effective index.\footnote{The analysis below shows that the relation between $\neff$ and $\omega$, as defined here, is independent of the gravitational field. Such notation and nomenclature thus facilitates comparison with established literature on fiber optics.}
Independent of heuristics based on geometrical optics, the given ansatz for $\psi$ yields a particularly simple form of the perturbative equations satisfied by $\t\ema{_b}$. In particular, matching terms of equal order in $\varepsilon$ and equal Fourier index, the equations reduce to a recursive system of the form
\begin{align}
	\label{eq:recursive system abstract}
	\OpH{m} \t*\ema{^{(j,m)}}
		&= \t*\Sigma{^{(j,m)}_\bulk}
		\,,
	&
	\OpD{m} \t*\ema{^{(j,m)}}
		&= \t*\Sigma{^{(j,m)}_\intf}
		\,,
\end{align}
where $\OpH{m}$ is a Helmholtz operator and $\OpD{m}$ is an operator encoding the discontinuities of the fields at the core-cladding interface, see \cref{app:Helmholtz Green,app:Junction} for explicit expressions.

\subsection{Homogeneous Equations}
\label{s:problem:homogeneous}

To understand the structure of the problem, it is instructive to consider, first, the homogeneous problem
\begin{align}
	\label{eq:problem:homogeneous}
	\OpH{m} \ema &= 0\,,
	&
	\OpD{m} \ema &= 0\,.
\end{align}
The general solution to $\OpH{m} a = 0$ that is regular on the baseline and decays exponentially in the cladding is given by
\begin{align}
	\label{eq:solution:homogeneous}
	f^{(m)}[\boldsymbol q]
		= (
			f_m(\boldsymbol q_0),
			f_m(\boldsymbol q_\parallel),
			f_{m+1}(\boldsymbol q_\sharp),
			f_{m-1}(\boldsymbol q_\flat)
		)\,,
\end{align}
where $\boldsymbol q_b = (q^\core_b, q^\clad_b)$ are coefficients that may depend on $\varsigma$, and $f_m$ is given explicitly in terms of Bessel functions:
\begin{align}
	f_m(q^\core, q^\clad)
		&= \begin{cases}
			q^\core \,J_m(u r) / J_m(u)	& \text{in the core},\\
			q^\clad K_m(w r) / K_m(w)	& \text{in the cladding}.
		\end{cases}
\end{align}
Here, the dimensionless coefficients $u$ and $w$ are given by
\begin{align}
	\label{eq:def u w}
	u
		&= (\ell_0 \freq / c) \ssqrt{n_1^2 - \neff^2}\,,
	&
	w
		&= (\ell_0 \freq / c) \ssqrt{\neff^2 - n_2^2}\,.
\end{align}
The interface condition $\OpD{m} a = 0$ then takes the form
\begin{align}
	\label{eq:interface M q}
	\OpD{m} f^{(m)}[\boldsymbol q]
		&\equiv \hOpD{m} \cdot \boldsymbol q
		= 0\,,
\end{align}
where $\hOpD{m}$ is a complex $8 \times 8$ matrix, see \cref{eq:junction matrix} for an explicit formula.
Non-trivial solutions to \cref{eq:problem:homogeneous} exist if and only if $\hOpD{m}$ is singular, i.e., $\det \hOpD{m} = 0$. As shown in \cref{app:Junction}, the determinant factorizes as
\begin{align}
	\det \hOpD{m}
		&\propto \t{\mathcal D}{_{\text{ph,}}_m} \t*{\mathcal D}{_{\text{g,}}^2_m}\,,
\end{align}
where roots of $\t{\mathcal D}{_{\text{ph,}}_m}$ give rise to physical modes (satisfying $\emGauge$ = 0 and $\t\emF{_\mu_\nu} \neq 0$), whereas $\t*{\mathcal D}{_{\text{g,}}_m} = 0$ yields so-called gauge modes ($\t\emF{_\mu_\nu} = 0$) and ghost modes ($\emGauge \neq 0$) that are irrelevant for the present consideration \cite{2022PhRvA.106f3511M,2023Mieling}.
The function $\t{\mathcal D}{_{\text{ph,}}_m}$ has the structure $\mathcal D_\text{ph}(|m|, v, \neff, n_1, n_2, \varrho/\ell_0)$, where $v$ is the normalized frequency $v = (\ell_0 \freq / c) \ssqrt{n_1^2 - n_2^2}$.
In a given fiber, non-trivial physical solutions to the homogeneous problem \eqref{eq:problem:homogeneous} exist if and only if $\neff$ is a root of $\t{\mathcal D}{_{\text{ph,}}_m}$ along the entire baseline, which entails the functional dependence $\neff = \neff(|m|, \omega)$.
Note that $\neff$ generally varies along the fiber due to its dependence on $\freq = \freqK / \lapse(s)$, but there is no explicit arc-length dependence of $\neff$.

If $\t{\mathcal D}{_{\text{ph}}_,_m} = 0$, the matrix $\hOpD{m}$ has a one-dimensional kernel and co-kernel, so there is an eight-component column vector $\boldsymbol{\hat q}{^m}$ and an eight-component row vector $\boldsymbol{\hat\kappa}{^m}$ satisfying
\begin{align}
	\label{eq:reference vectors}
	\hOpD{m} \cdot \boldsymbol{\hat q}{^m} &= 0\,,
	&
	\boldsymbol{\hat\kappa}{^m} \cdot \hOpD{m} &= 0\,,
\end{align}
respectively, and every other solution $\boldsymbol{q}{^m}$ (or $\boldsymbol{\kappa}{^m}$) to such an equation is a multiple of the reference solution $\boldsymbol{\hat q}{^m}$ (or $\boldsymbol{\hat\kappa}{^m}$), where the coefficient of proportionality may depend on $\varsigma$.

\subsection{Inhomogeneous Equations}
\label{s:problem:inhomogeneous}

Whereas the leading-order terms in \cref{eq:ansatz a} satisfy homogeneous equations as given in \cref{eq:problem:homogeneous}, higher-order terms satisfy inhomogeneous equations of the form
\begin{align}
	\label{eq:problem:inhomogeneous}
	\OpH{m} a &= \t*\Sigma{_\bulk}\,,
	&
	\OpD{m} a &= \t*\Sigma{_\intf}\,.
\end{align}
The general solution to the inhomogeneous Helmholtz equation is given by
\begin{align}
	a
		&= f^{(m)}[\boldsymbol q] + \OpG{m} \t*\Sigma{_\bulk}\,,
\end{align}
where $f^{(m)}[\boldsymbol q]$ is defined in \cref{eq:solution:homogeneous} and $\OpG{m}$ is the Green operator defined in \cref{app:Helmholtz Green}.
The interface condition in \cref{eq:problem:inhomogeneous} then takes the form
\begin{align}
	\label{eq:inhomogeneous matching q}
	\hOpD{m} \cdot \boldsymbol q
		+ \OpD{m} \OpG{m} \t\Sigma{_\bulk}
		&= \t\Sigma{_\intf}\,,
\end{align}
where $\hOpD{m}$ is the same matrix as in \cref{eq:interface M q}.
Now, if $\hOpD{m}$ is non-singular, this equation admits a unique solution for $\boldsymbol q$. This non-resonant case occurs whenever the homogeneous problem only admits trivial solutions. If, alternatively, $\hOpD{m}$ is singular (resonant case), a necessary and sufficient condition for solvability of this equation is
\begin{align}
	\label{eq:condition solvability}
	\boldsymbol{\hat\kappa}{^m} \cdot \OpD{m}(\OpG{m} \t*\Sigma{_\bulk} - \t*\Sigma{_\intf}) = 0\,,
\end{align}
where $\boldsymbol{\hat\kappa}{^m}$ spans the co-kernel of $\hOpD{m}$, see \cref{eq:reference vectors}.
If \cref{eq:condition solvability} is satisfied, \cref{eq:inhomogeneous matching q} determines the coefficients $\boldsymbol q$ up to a homogeneous solution that is proportional to $\boldsymbol{\hat q}{^m}$.

\subsection{Perturbative Equations}

Having established the main facts about the homogeneous equation~\eqref{eq:problem:homogeneous} and the inhomogeneous equation~\eqref{eq:problem:inhomogeneous}, the system~\eqref{eq:recursive system abstract} can be solved iteratively.
Specifically, separating the leading-order equations from the higher-order ones, one obtains
\begin{subequations}
\begin{align}
	\label{eq:recursive system start}
	\OpH{m} \t*\ema{^{(0,m)}}
		&= 0
		\,,
	&
	\OpD{m} \t*\ema{^{(0,m)}}
		&= 0
		\,,
	\\
	\label{eq:recursive system iteration}
	\OpH{m} \t*\ema{^{(j,m)}}
		&= \t*\Sigma{^{(j,m)}_\bulk}
		\,,
	&
	\OpD{m} \t*\ema{^{(j,m)}}
		&= \t*\Sigma{^{(j,m)}_\intf}
		\,,
\end{align}
\end{subequations}
with $m \in \mathbf Z$ and $j \geq 1$, where $\t*\Sigma{^{(j,m)}_\bulk}$ and $\t*\Sigma{^{(j,m)}_\intf}$ depend on the fields $\t*\ema{^{(j',m')}} $ of order $j' < j$.

As demonstrated in \cref{s:problem:homogeneous}, for each $m \in \mathbf Z$, the system \eqref{eq:recursive system start} admits a non-trivial physical solution if and only if $\t{\mathcal D}{_{\text{ph,}}_m} = 0$, which is an equation involving $\neff(\varsigma)$, $n_1$, $n_2$, $\freq(\varsigma)$, and the core radius $\varrho$.
These equations for different values of $|m|$ are mutually exclusive and can thus be solved only for at most two values of $m$ that differ in sign. The following analysis focuses on the single-mode regime for which one has $m = \pm 1$ (this case allows for a particularly simple description of the electromagnetic polarization).
For these “on-shell” Fourier indices $m = \pm 1$, the fields $\t*\ema{^{(0,m)}}$ are determined uniquely up to amplitude factors of the form $\t*{\mathcal A}{_\pm^{(0)}}(\varsigma)$, whereas all “off-shell” Fourier indices $m \neq \pm 1$ do not contribute at leading order: $\t*\ema{^{(0,m)}} = 0$.

At next order, $j = 1$, the analysis in \cref{s:problem:inhomogeneous} shows that the inhomogeneous equations with $m$ on-shell can be solved if and only if
\begin{align}
	\boldsymbol{\hat\kappa}{^m} \cdot \OpD{m}(\OpG{m} \t*\Sigma{_\bulk^{(1,m)}} - \t*\Sigma{_\intf^{(1,m)}}) = 0\,.
\end{align}
Since $\t*\Sigma{_\bulk^{(1,m)}}$ depends on $\t\p{_\varsigma} \t*\ema{^{(0,m)}}$, these constraints determine the variation of the amplitudes $\t*{\mathcal A}{_\pm^{(0)}}$ along $\varsigma$ (explicit equations are presented below).
\Cref{eq:recursive system iteration} with $m$ on-shell and $j=1$ can then be solved for $\t*\ema{^{(1,\pm)}}$ up to free amplitudes $\t*{\mathcal A}{_\pm^{(1)}}(\varsigma)$ whereas the equations with $m$ off-shell have unique solutions without such free amplitudes.

The scheme just described extends to all $j \geq 1$.
At each order, the solvability conditions
\begin{align}
	\label{eq:condition solvability at order j}
	\boldsymbol{\hat\kappa}{^m} \cdot \OpD{m}(\OpG{m} \t*\Sigma{_\bulk^{(j,m)}} - \t*\Sigma{_\intf^{(j,m)}}) = 0
\end{align}
with $m = \pm 1$ determine transport equations for the amplitudes $\t*{\mathcal A}{_\pm^{(j-1)}}(\varsigma)$ that are not determined from the previous-order equations.
With \cref{eq:condition solvability at order j} satisfied, \cref{eq:recursive system iteration} with $m$ on-shell can be solved uniquely up to two amplitudes $\t*{\mathcal A}{_\pm^{(j)}}(\varsigma)$ (to be determined at next order), and \cref{eq:recursive system iteration} with $m$ off-shell can be solved uniquely without any free parameters.

The terms arising in the second-order expansion of the metric tensor that are given explicitly in \cref{sec:setup} determine the solvability conditions \eqref{eq:condition solvability at order j} for $j \in \{0, 1, 2\}$.
As shown in \cref{app:transport equations}, these equations can be cast in the form of transport equations for
the phase perturbation $\delta \psi = \pm \int \delta \beta \dd s$, where $\delta \beta = \beta - \neff \freq /c$ accounts for all perturbation terms in \cref{eq:ansatz beta} that contribute to the phase $\psi$ in \cref{eq:ansatz phase},
and the Jones vector $\t\jones{_\alpha}$ (a complex transverse unit vector describing the state of polarization)
along the baseline $\baseline$:
\begin{subequations}
\begin{align}
	\label{eq:transport phase perturbation}
	\DD \delta \psi
		&= \delta \beta
		+ O(\varepsilon^3/\ell_0)
		\,,
	\\
	\label{eq:transport polarization}
	\DD \t\jones{_\alpha}
		&= \varpi \t\twist{_\alpha^\beta} \t \jones{_\beta}
		+ \i \t H{_\alpha^\beta} \t\jones{_\beta}
		+ O(\varepsilon^3/\ell_0)
		\,.
\end{align}
\end{subequations}
Here, $\DD$ is the spatial covariant derivative $\DD = \pm \t\del{_s}$ with the sign depending on the direction of light propagation as in \cref{eq:ansatz phase}.
Explicitly, the perturbation of the propagation coefficient, $\delta \beta$, in \cref{eq:transport phase perturbation} can be written as
\begin{align}
	\label{eq:perturbation beta}
	\begin{split}
		\delta\beta
			={}&
			  \t\gamma{_1} \t\metric{^\alpha^\beta} \t z{_\alpha} \t z{_\beta}
			+ \t\gamma{_2} (K  + \t\metric{^\alpha^\beta} \t\lapse{_,_\alpha_\beta} / \lapse)
			\\&
			+ \t\gamma{_3} \t\riemann{_s_s}
			+ \t\gamma{_4} \t\p{_s}\lapse/\lapse
			+ \t\gamma{_5} \t\p{_s} \t\p{_s}\lapse/\lapse
			\,,
	\end{split}
\end{align}
where $\t\gamma{_J}$ are dimensionful coupling coefficients (phase coupling moments),
$K = \t\riemann{_x_y_x_y}$ is the sectional curvature of the transverse plane,
$\t\riemann{_s_s}$ is the longitudinal projection of the spatial Ricci tensor $\t\riemann{_i_j}$,
and the vector $\t z{_\alpha} = \t\pnormal{_\alpha} + \t\lapse{_,_\alpha} / \lapse$ combines the effects of the principal normal $\t\pnormal{_\alpha}$ and logarithmic derivatives of the lapse function $\lapse$ in transverse directions.
The dimensionless coefficient $\varpi$ in \cref{eq:transport polarization} describes the coupling to the antisymmetric tensor $\t\twist{_\alpha_\beta} = \lapse \t{(\d \shift)}{_\alpha_\beta}$,
and $\t H{_\alpha_\beta}$ is a symmetric and trace-free tensor of the form
\begin{align}
	\label{eq:Spin Hall tensor}
	\begin{split}
		\t H{_\alpha_\beta}
			={}&
			\t\eta{_1} (\t z{_\alpha} \t z{_\beta} - \half \t\metric{_\alpha_\beta} \t\metric{^\gamma^\delta} \t z{_\gamma} \t z{_\delta})
			\\&
			+ \t\eta{_2} (\t w{_\alpha_\beta} - \half \t\metric{_\alpha_\beta} \t\metric{^\gamma^\delta} \t w{_\gamma_\delta})\,.
	\end{split}
\end{align}
Similarly to the coefficients $\t\gamma{_J}$ in \cref{eq:perturbation beta}, $\t\eta{_K}$ are dimensionful coupling coefficients (polarization coupling moments), and $\t w{_\alpha_\beta} = \t\Weyl{_\alpha_s_\beta_s}$ is a projection of the space-time Weyl tensor $\t\Weyl{_\mu_\nu_\rho_\sigma}$.

The explicit form of these transport equations can be derived using computer algebra systems – a reference implementation written in \textit{Wolfram Language} is available online \cite{mieling_2024_14142452}.
The values of the coupling coefficients $\varpi$, $\t\gamma{_J}$, and $\t\eta{_K}$ depend on the normalized frequency $v$ and must generally be determined by quadrature (see the next section for numerical results).

\section{Results}

\begin{figure*}[t]
	\begin{minipage}[t]{\columnwidth}
		\vspace{0\baselineskip}
		\includegraphics{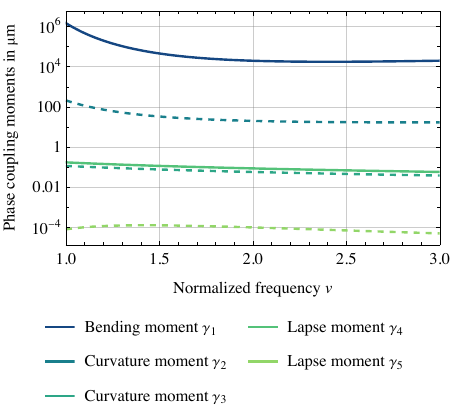}
	\end{minipage}
	\hfill
	\begin{minipage}[t]{\columnwidth}%
		\vspace{0\baselineskip}
		\includegraphics{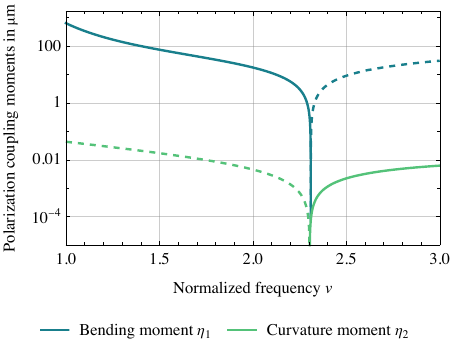}
	\end{minipage}
	\caption{%
		Dependence of the phase coupling moments $\t\gamma{_J}$ (left) and polarization coupling moments $\t\eta{_K}$ (right) on the normalized frequency $v = (\varrho \freq/c) \ssqrt{n_1^2 - n_2^2}$, where $\varrho$ is the fiber’s core radius, $\freq$ is the optical frequency, and $n_1$ and $n_2$ are the refractive indices in the core and cladding, respectively.
		In these logarithmic plots, solid lines indicate positive values whereas dashed lines correspond to negative values.
		The values shown here were obtained for a single-mode fiber with core radius $\varrho = \SI{4.1}{\micro\meter}$ and refractive indices $n_1 = 1.4712$ and $n_2 = 1.4659$.
	}
	\label{fig:coupling moments}
\end{figure*}

The results derived above can be summarized as a series of transport equations of the physical frequency $\freq$, effective index $\neff$, phase $\psi$, and transverse Jones vector $\t\jones{^i}$ along the direction of light propagation:
\begin{subequations}
\label{eq:summary:transport equations}
\begin{align}
	\label{eq:summary:transport frequency}
	\DD \freq =&
		- \freq\, \DD \ln \lapse\,,
	\\
	\label{eq:summary:transport index}
	\DD \neff =&
		- \freq \frac{\d \neff}{\d \freq} \DD \ln \lapse\,,
	\\
	\label{eq:summary:transport phase}
	\DD \psi =&
		+ \freq \neff / c
		\pm \freq \lapse \t\shift{_s} / c
		+ \delta \beta
		+ O(\varepsilon^3 / \ell_0)\,,
		\\
	\begin{split}
		\label{eq:summary:transport polarization}
		\DD \t\jones{^i} =&
			\pm \curvature \t\epsilon{^i_j_k} \t b{^j} \t\jones{^k}
			+ \varpi \t\epsilon{^i_j_k} \t\gyrotropy{^j} \t\jones{^k}
			\\&
			+ \i \t H{^i_k} \t\jones{^k}
			+ O(\varepsilon^3 / \ell_0)\,,
	\end{split}
	\end{align}
\end{subequations}
where the upper and lower signs apply to light propagating in parallel or antiparallel to the orientation of the fiber’s baseline $\baseline$, respectively.
In these equations, $\DD$ denotes the covariant derivative along the direction of light propagation, i.e., $\DD = \pm \t\utangent{^i} \t\del{_i}$, where $\t\utangent{^i}$ denotes the unit tangent of $\baseline$.
The quantities $\curvature$ and $\t\ubinormal{^i}$ denote, respectively, the associated curvature and unit binormal. Apart from the spatial geometry of the fiber, these equations also depend on the space-time geometry through the lapse $\lapse$ and the shift $\t\shift{_i}$ (both evaluated on the baseline).
The effective index $\neff$ depends on the frequency $\freq$ (as defined relative to the fiber’s rest frame) according to a dispersion relation that takes the same form as for straight fibers in flat space-time (see Ref.~\cite[Sect.~3.1]{2005Liu} for details).
The term $\delta \beta$ in \cref{eq:summary:transport phase} constitutes a perturbation to the propagation parameter and is given explicitly in \cref{eq:perturbation beta}.
The first term in \cref{eq:summary:transport polarization} accounts for the difference between parallel transport and Fermi–Walker transport, see \cref{eq:Fermi Walker}. The second term in \cref{eq:summary:transport polarization} describes a rotation of the electromagnetic polarization about the vector
\begin{align}
	\label{eq:gyrotropy vector}
	\t\gyrotropy{^i}
		&= \half \t t{^i} \t\epsilon{^j^k^l} \t t{_j} \t{\twist}{_k_l}
		\equiv \lapse\, \t t{^i} \t\epsilon{^j^k^l} \t t{_j} \t\p{_k} \t\shift{_l}\,,
\end{align}
with a rotation rate that depends on the dimensionless coupling constant $\varpi$.
Finally, the last term in \cref{eq:summary:transport polarization} gives rise to additional polarization dynamics that are described by the Hermitian, transverse, and traceless tensor given in \cref{eq:Spin Hall tensor}.

Whereas the dimensionful phase and polarization moments depend strongly on the normalized frequency $v$, see \cref{fig:coupling moments}, variations of the gyrotropy factor $\varpi$ are much smaller. 
For example, in fibers with refractive indices $n_1 = 1.4712$ and $n_2 = 1.4659$ (as were considered in Ref.~\cite{2018CQGra..35x4001B}) and normalized frequencies $v = (\varrho \freq / c) \ssqrt{n_1^2 - n_2^2}$ in the range $1 < v < 3$, the coefficients $\t\gamma{_J}$ and $\t\eta{_K}$ vary over multiple orders of magnitude while numerical variations of $\varpi \approx 0.341$ are on the order of \SI{0.2}{\percent}.
Hence, the gyrotropy factor $\varpi$ for any given fiber can be considered as constant with negligible error.

The following paragraphs elaborate on details of the transport equations \eqref{eq:summary:transport frequency} to \eqref{eq:summary:transport polarization}.

\paragraph{Optical Phase and Redshift}

The gravitational redshift along the fiber is described by \cref{eq:summary:transport frequency}, which has the exact solution $\freq(s) / \freq(0) = \lapse(0) / \lapse(s)$.
Even in the absence of material dispersion, waveguide dispersion entails that the redshift induces a change in effective index $\neff$ along the fiber according to \cref{eq:summary:transport index}.
According to \cref{eq:summary:transport phase}, the integral $\int \neff \omega \dd s / c$ then determines the leading-order contribution to the phase $\psi$.
If the lapse $\lapse$ is constant along the fiber, this leading-order contribution reduces to the standard form $\neff \omega s / c$, but more general cases do not admit closed-form solutions.
Whereas previously published derivations of such effects in \SMF\ were restricted to specially aligned fibers \cite{2018CQGra..35x4001B,2022PhRvA.106f3511M}, the present formulas are free of such restrictions and hold also for fiber spools as will be used in the forthcoming \GRAVITES\ experiment \cite{2017NJPh...19c3028H,2024Polini}.
While the interplay between the redshift and dispersion is beyond the current experimental sensitivity of the Mach–Zehnder interferometers used \cite[Sect.~4.7]{2018CQGra..35x4001B}, dispersive effects give rise to alternative methods for measuring the gravitational redshift in fibers \cite{2001PhRvD..63f2003M}.

\paragraph{Sagnac Effect}

The second term in \cref{eq:summary:transport phase} accounts for phases that are induced by the shift $\t\shift{_i}$. As this term changes sign when reversing the direction of light propagation, it can be isolated by interfering two waves that propagate in opposite directions through a fiber loop to produce the phase shift $\Delta \psi =  2 \freqK \oint_\baseline \t\shift{_i} \t{\d x}{^i} / c$, where $\freqK = \freq \lapse$ denotes the constant Killing frequency.%
\footnote{%
	Since the Sagnac phase depends on the shift $\t\shift{_i}$ through the loop integral $\oint \t\shift{_i} \t{\d x}{^i}$, it is invariant under the gauge transformations described in footnote~\ref{ft:gauge transform shift}. For an interpretation of the Sagnac phase in terms of the twist of the Killing vector field $\t\Killing{^\mu}$, see Ref.~\cite{1975JMP....16..341A}.%
}
The present derivation of the Sagnac phase provides a rigorous extension of standard geometric-optics derivations \cite{1967RvMP...39..475P,1975JMP....16..341A} to fiber optics without assuming specific fiber alignment as was done in previous analyses \cite{1982ApOpt..21.1400L,2020CQGra..37v5001M}.
Apart from Sagnac effects arising in interferometers rotating relative to Earth and those arising from Earth’s spin \cite{2013OptFT..19..828L,2023PhRvR...5b2005C,2024SciA...10O.215S}, the above derivation also applies to frame-dragging-induced Sagnac phases of the kind described in Ref.~\cite{2022AVSQS...4a1401K}.

\paragraph{Higher-Order Phase Corrections}

The last term in \cref{eq:summary:transport phase}, with $\delta\beta$ given explicitly in \cref{eq:perturbation beta}, describes higher-order phase corrections.
They depend on various components of the space-time curvature tensor $\t{\Riemann}{_\mu_\nu_\rho_\sigma}$ such as $\t{\Riemann}{_t_\alpha_t_\beta} = \lapse \t\del{_\alpha} \t\del{_\beta} \lapse$ and $\t{\Riemann}{_s_\alpha_s_\beta} = \t\riemann{_s_\alpha_s_\beta}$, where $\lapse$ is the lapse function and $\t\riemann{_i_j_k_l}$ is the spatial curvature tensor.
As can be seen in \cref{fig:coupling moments}, however, the dominant effect arises from the first term in \cref{eq:perturbation beta} that depends on the vector $\t z{_\alpha} = \t\pnormal{_\alpha} + \t\del{_\alpha} \lapse / \lapse$. Since, the curvature of optical fiber spools [on the order of $1/(\SI{0.1}{\meter})$] is significantly larger than the corresponding term arising from gravitational acceleration [$\mathrm g/c^2 \approx 1/(\SI{9e15}{\meter})$], bending effects as computed in Ref.~\cite{2023PhRvR...5b3140M} far exceed effects arising from non-trivial space-time geometry.

\paragraph{Rytov’s Law}

The leading-order term in \cref{eq:summary:transport polarization}, $\pm \curvature \t\epsilon{^i_j_k} \t b{^j} \t\jones{^k}$, ensures that $\t\jones{^i}$ remains orthogonal to the direction of light propagation.
If higher-order terms are negligible, the transport law for the polarization vector $\t\jones{_i}$ reduces to Fermi–Walker transport, see \cref{eq:Fermi Walker}.
At this level of approximation, $\t\jones{_i}$ rotates relative to the transverse Frenet–Serret basis $(\t\unormal{^i}, \t\ubinormal{^i})$ according to $\d\phi / \d s = - \torsion$, where $\torsion$ is the torsion of the baseline, see \cref{eq:Frenet Serret}.
This phenomenon is known as Rytov’s law and has been demonstrated experimentally \cite{1984OQEle..16..455R,1986PhRvL..57..937T}.
Whereas previous derivations of this effect were restricted to flat space-time \cite{1926Bortolotti,1938Rytov_\transliteration,1987Natur.326..277B,2018PhRvA..97c3843L,2019PhRvA.100c3825L,2023PhRvR...5b3140M}, the analysis here shows how Rytov’s law extends to curved space.

\paragraph{Geometrical Gyrotropy}

The second term in \cref{eq:summary:transport polarization} describes an additional rotation in the transverse plane whose magnitude equals the norm of $\t\gyrotropy{^i}$, as defined in \cref{eq:gyrotropy vector}, multiplied by the coupling constant $\varpi$.
Such a polarization rotation was previously computed for light passing a rotating mass distribution using ray-optics approximations \cite{1957Skrotskii_\transliteration,1960PhRv..118.1396P}. This effect is commonly referred to as the gravitational Faraday effect owing to its mathematical analogy to the Faraday effect in media under the influence of external magnetic fields \cite{1999PhRvD..60b4013N}.
However, as the shift-induced rotation is reciprocal while the standard Faraday effect is non-reciprocal \cite[Sect.~3.7.4]{2004Damask}, such an effect can more accurately be regarded as a reciprocal gyrotropy effect that is analogous to optical activity in chiral materials.
Since the present calculation is indifferent to the origin of $\t\shift{_i}$, it also applies to rotating systems in flat space-time.
In particular, for an \SMF\ with refractive indices as in \cref{fig:coupling moments}, the gyrotropy effect induced by a rotational frequency $f$ causes a polarization rotation on the order of $(\SI{e-8}{\radian\per\meter}) \cdot (f/\si{\hertz})$.

\paragraph{Inverse Spin Hall Effects}

The tensor $\t H{_i_j}$ entering \cref{eq:summary:transport polarization} describes inverse spin Hall effects.
Whereas the ordinary spin Hall effect of light describes how the polarization affects the trajectories of light rays \cite{2008Sci...319..787H,2008NaPho...2..748B,2020PhRvD.102b4075O}, in optical fibers the light trajectory is constrained and, instead, the polarization vector undergoes non-trivial dynamics, giving rise to an inverse spin Hall effect \cite{2023PhRvR...5b3140M}.
In the present case, \cref{eq:Spin Hall tensor} shows that inverse spin Hall effects arise from fiber bending (described by the principal normal $\t\pnormal{_\alpha}$), transverse gravitational acceleration (arising from logarithmic derivatives of the lapse function, $\t\del{_\alpha}\lapse/\lapse$), and the space-time Weyl tensor $\t\Weyl{_\mu_\nu_\rho_\sigma}$.
The size of the respective coupling coefficients are plotted in \cref{fig:coupling moments}: similarly to the higher-order phase corrections, effects arising from gravitational acceleration and space-time curvature are multiple orders of magnitude smaller than those arising from fiber bending.
Explicit expressions for such effects in helical fibers can be found in Ref.~\cite{2023PhRvR...5b3140M}.

In conclusion, the transport equations \eqref{eq:summary:transport frequency} to \eqref{eq:summary:transport polarization} provide a generalization of multiple partial results that were derived previously.
Specifically, prior models of the redshift and Sagnac effect in optical fibers were limited in admissible fiber geometries and considered these effects in isolation \cite{2018CQGra..35x4001B,2022PhRvA.106f3511M,2023CQGra..40n5008S,2024NJPh...26h3010B,1981OptL....6..401A,2020CQGra..37v5001M}.
The prior results on the redshift are obtained from \cref{eq:summary:transport equations} by keeping only the $\lapse$-terms (in particular, setting $\t\shift{_i} = 0$), and those on the Sagnac effect are reproduced by keeping only the $\t\shift{_i}$-terms (and setting $\lapse = 1$).
Their unified description in general fiber geometries, however, constitutes a new result.
Similarly, by taking $\lapse = 1$, $\t\shift{_i} = 0$ and $\t\riemann{_i_j_k_l} = 0$, \cref{eq:summary:transport polarization} reproduces the results of Ref.~\cite{2023PhRvR...5b3140M} that derived Rytov’s law and inverse spin Hall effects for bent fibers in inertial systems.
In this sense, the present model gives a unified description of bent fibers in non-inertial systems.
Additionally, \cref{eq:summary:transport equations} contains terms that depend on the space-time curvature tensor – these expressions thus cannot be reproduced by considering non-inertial motion in flat space-time. While some of the these terms are analogous to expressions arising in curvature-couplings in free-space light propagation, their explicit form in fiber optics constitutes a new result of the perturbation theory developed here.

\section{Discussion}

Generally speaking, gravitational effects on light propagation in media can be categorized into direct and indirect effects.
Whereas direct effects arise from the dependence of Maxwell’s equations on the space-time metric (more precisely, from the explicit dependence of the constitutive equation on $\t\Metric{_\mu_\nu}$), indirect effects arise from the medium’s response to the gravitational field.
The calculations provided here are restricted to direct effects as the cross-section of the fiber was assumed to be translation-invariant and the medium was modeled as linear, homogeneous, isotropic, and non-dispersive with constant refractive indices.
The present analysis shows that the direct influence of gravity on light propagation in optical fibers is not limited to the gravitational redshift (previous calculations of which were restricted to specific fiber alignments and linearized gravity), but also includes higher-order phase perturbations and inverse spin Hall effects that depend on the curvature tensor of space-time.

Some indirect effects of gravity on light propagating in optical fibers have already been analyzed in the literature. Dispersion arising in modes of constant Killing frequency due to the gravitational redshift are described in Refs.~\cite{2001PhRvD..63f2003M,2018CQGra..35x4001B}, geometric deformation effects and photoelastic effects in horizontally aligned fibers are modeled in Ref.~\cite{2024OpenResEuro...BCS}, and similar photoelastic effects on solitons in vertically suspended fibers are analyzed in Refs.~\cite{2023CQGra..40n5008S,2024NJPh...26h3010B}.
The perturbative framework developed here could be used to extend these results to bent fibers and more general setups.

In addition to gravitational effects described above, the calculations presented here also account for a shift $\t\shift{_i}$ that need not be of gravitational origin (shift vectors arise in rotating reference systems even if the ambient space-time is flat).
While shift vectors are known to induce Sagnac phase shifts, explicit calculations of this effect were, so far, limited to ray-optics approximations or wave-optics calculations in specific fiber geometries.
The present analysis provides a rigorous extension of these results to bent single-mode fibers without using ray approximations.
Additionally, the calculation shows that light polarization undergoes a rotation whenever the shift vector has a non-zero curl.
This geometrical gyrotropy effect is analogous to the gravitational Faraday effect.

Overall, the geometrical description of optical fibers developed here, specifically the use of the spatial metric $\t\metric{_i_j}$ in \cref{eq:metric general} rather than the \ADM\ metric $\t*g{^\ADM_i_j}$ and the geometrically defined coordinate system based on Bishop’s frame satisfying \cref{eq:frame Bishop transport}, results in covariant transport laws \eqref{eq:summary:transport frequency} to \eqref{eq:summary:transport polarization} for the optical phase and polarization.
This general scheme could be combined with numerical methods to describe light-propagation in optical fibers with other cross-sections (such as \SMF\ that deviate from perfect translational symmetry along the axis or from rotational symmetry in the transverse planes, but also polarization-maintaining fibers or hollow-core fibers with more complex internal structures), as well as dispersive and non-linear media (as is relevant to the description of optical solitons).

\section*{Acknowledgments}
We are grateful to Marius Oancea and Piotr Chruściel for helpful discussions.
Research by T.B.M.\ is funded by the European Union (\textsc{erc}, \textsc{gravites}, project no.~101071779).
Views and opinions expressed are however those of the authors only and do not necessarily reflect those of the European Union or the European Research Council Executive Agency. Neither the European Union nor the granting authority can be held responsible for them.
Research by M.H.\ was supported by an MSc Fellowship from the Vienna Doctoral School in Physics (\textsc{vdsp}).

\appendix

\section{Notation and Conventions}
\label[appendix]{app:notation}

Throughout, this paper uses the Einstein summation convention for repeated indices (that applies whenever an index is printed once as a subscript and once as a superscript, but not to repeated subscripts or repeated superscripts). In these sums, the indices $\mu$, $\nu$, \ldots range from $0$ to $3$; $i, j, \ldots$ from $1$ to $3$; $\alpha$, $\beta$, \ldots from $1$ to $2$; $A$, $B$, \ldots range over the symbols $+$ and $-$; and frame indices $a$, $b$, \ldots range over the index set $\{0, \parallel, \sharp, \flat\}$.

The sign conventions used here are the same as in Ref.~\cite{1973MTW}.
Specifically, the signature of the metric $\t\Metric{_\mu_\nu}$ is $(-, +, +, +)$, the Riemann tensor $\t{{\Riemann}}{^\mu_\nu_\rho_\sigma}$ satisfies the Ricci identity in the form $\t\nabla{_\rho} \t\nabla{_\sigma} \t X{^\mu} - \t\nabla{_\sigma} \t\nabla{_\rho} \t X{^\mu} = \t{{\Riemann}}{^\mu_\nu_\rho_\sigma} \t X{^\nu}$, and the Ricci tensor is defined as $\t{{\Riemann}}{_\mu_\nu} = \t{{\Riemann}}{^\rho_\mu_\rho_\nu}$.

The exterior derivative of a differential form $\t\alpha{_\nu_\rho_\ldots}$ is denoted by $\t{(\d \alpha)}{_\mu_\nu_\rho_\ldots}$, and the divergence of a tensor density $\t{\mathfrak T}{^\mu^\nu^\rho^\ldots}$ is denoted by $\t{(\div \mathfrak T)}{^\nu^\rho^\ldots}$.
Symmetrization and anti-symmetrization of indices is indicated by parentheses and brackets, respectively.

\section{Helmholtz Operators and Green Operators}
\label[appendix]{app:Helmholtz Green}

The Helmholtz operator entering \cref{eq:recursive system abstract} acts on a set of fields $\ema = (\t\ema{_0}, \t\ema{_\parallel}, \t\ema{_\sharp}, \t\ema{_\flat})$ as
\begin{align}
	\label{eq:Helmholtz operator}
	\OpH{m} \ema
		&= (
			\opH{m} \t\ema{_0},
			\opH{m} \t\ema{_\parallel},
			\opH{m+1} \t\ema{_\sharp},
			\opH{m-1} \t\ema{_\flat}
		)\,,
	\\
	\shortintertext{with}
	\opH{m}
		&= \frac{\p^2}{\p r^2}
		+ \frac{1}{r} \frac{\p}{\p r}
		- \frac{m^2}{r^2}
		+ (\ell_0^2 \freq^2/c^2) (n^2 - \neff^2)
		\,.
\end{align}
Here, $n$ is the local refractive index taking the constant values $ n_1$ in the core, and $n_2$ in the cladding.
This operator is known from the study of fiber optics in flat space-time \cite[Sect.~III.A.]{2022PhRvA.106f3511M}

The associated Green operator $\OpG{m}$ that is used in \cref{s:problem:inhomogeneous} is defined as
\begin{align}
	\label{eq:Green operator}
	\OpG{m} \ema
		&= (
			\opG{m} \t\ema{_0},
			\opG{m} \t\ema{_\parallel},
			\opG{m+1} \t\ema{_\sharp},
			\opG{m-1} \t\ema{_\flat}
		)\,,
	\\\shortintertext{with}
	\opG{m} f
		&= \begin{cases}
			\opGcore{m} f & \text{in the core},\\
			\opGclad{m} f & \text{in the cladding},
		\end{cases}
\end{align}
in which $\opGcore{m}$ and $\opGclad{m}$ are defined as
\begin{subequations}
\begin{align}
	\label{eq:operator Green core}
	\begin{split}
		\opGcore{m} f(r)
			={}& +\frac{\pi}{2} Y_m(u r) \int_0^r \hspace{-0.5em} J_m(u \rho) f(\rho) \rho \dd \rho
			\\&
			+ \frac{\pi}{2} J_m(u r) \int_r^{r_*} \hspace{-0.5em} Y_m(u \rho) f(\rho) \rho \dd \rho\,,
	\end{split}
	\\
	\label{eq:operator Green cladding}
	\begin{split}
		\opGclad{m} f(r)
			={}& - I_m(w r) \int_r^\infty \hspace{-0.5em} K_m(w \rho) f(\rho) \rho \dd \rho
			\\&
			- K_m(w r) \int_{r_*}^r \hspace{-0.5em} I_m(w \rho) f(\rho) \rho \dd \rho\,.
	\end{split}
\end{align}
\end{subequations}
The free parameter $r_*$ can be chosen arbitrarily to simplify explicit calculations by making the second terms in \cref{eq:operator Green core,eq:operator Green cladding} vanish at any fixed radius.
Independent of the choice of $r_*$, this operator satisfies $\OpH{m} \OpG{m} a = a$, as can be verified using the Wroński determinants
\begin{align}
	\begin{vmatrix}
			J_m(x)
		&	Y_m(x)
		\\
			J_m'(x)
		&	Y_m'(x)
	\end{vmatrix}
	&= \frac{2}{\pi x}\,,
	&
	\begin{vmatrix}
			K_m(x)
		&	I_m(x)
		\\
			K_m'(x)
		&	I_m'(x)
	\end{vmatrix}
	&= \frac{1}{x}\,,
\end{align}
see, e.g., eqs.~(9.1.6) and (9.6.15) in Ref.~\cite{1972Olver}.

\section{Interface Conditions}
\label[appendix]{app:Junction}

The explicit form of the discontinuity operator $\OpD{m}$ that arises in \cref{eq:recursive system abstract} is given by
\begin{align}
	\OpD{m} \ema
		&= \begin{pmatrix}
				\jump{\t\ema{_0}}
			\\	\jump{\t\ema{_\parallel}}
			\\	\jump{\t\ema{_\sharp}}
			\\	\jump{\t\ema{_\flat}}
			\\	\jump{\t\p{_r} \t\ema{_\parallel}}
			\\	\jump{\opC{m+1}{-} \t\ema{_\sharp} - \opC{m-1}{+} \t\ema{_\flat}}
			\\	\jump{\i n^2 \freq \t\ema{_0} + \opC{m+1}{-} \t\ema{_\sharp} + \opC{m-1}{+} \t\ema{_\flat}}
			\\	\jump{n^2\{ \t\p{_r} \t\ema{_0} + \tfrac{\i \freq}{\sqrt 2} (\t\ema{_\sharp} + \t\ema{_\flat}) \}}
		\end{pmatrix}\,,
\end{align}
where $\jump{f} = (f|_{r \downarrow (\varrho/\ell_0)}) - (f|_{r \uparrow (\varrho/\ell_0)})$ denotes the jump of a function at the core-cladding interface, and $\opC{m}{\pm} = \tfrac{1}{\sqrt2} (\t\p{_r} \mp m/r)$.

The corresponding matrix $\hOpD{m}$ as defined in \cref{eq:interface M q} has the explicit form
\begin{widetext}
	\begin{align}
		\label{eq:junction matrix}
		\hOpD{m}
		=
		\begin{pmatrix*}[c]
			+1 & 0 & 0 & 0 & -1 & 0 & 0 & 0 \\
			0 & +1 & 0 & 0 & 0 & -1 & 0 & 0 \\
			0 & 0 & -u \t*{\mathscr J}{_m^+} & 0 & 0 & 0 & +w \t*{\mathscr K}{_m^+} & 0 \\
			0 & 0 & 0 & +u\t*{\mathscr J}{_m^-} & 0 & 0 & 0 & +w \t*{\mathscr K}{_m^-} \\
			0 & + u^2 \t*{\mathscr J}{_m} & 0 & 0 & 0 & - w^2 \t*{\mathscr K}{_m} & 0 & 0 \\
			0 & 0 & +u & +u & 0 & 0 & +w & -w
			\\
				+\i n_1^2 \freq
			&	0 & +\frac{u}{\sqrt{2}}
			&	-\frac{u}{\sqrt{2}}
			&	-\i n_2^2 \freq
			&	0
			&	+\frac{w}{\sqrt{2}}
			&	+\frac{w}{\sqrt{2}}
			\\
				+ n_1^2 u^2 \t*{\mathscr J}{_m}
			&	0
			&	-\frac{\i n_1^2 u \freq}{\sqrt{2}} \t*{\mathscr J}{_m^+}
			&	+\frac{\i n_1^2 u \freq}{\sqrt{2}} \t*{\mathscr J}{_m^-}
			&	-n_2^2 w^2 \t*{\mathscr K}{_m}
			&	0
			&	+\frac{\i n_2^2 w \freq}{\sqrt{2}} \t*{\mathscr K}{_m^+}
			&	+\frac{\i n_2^2 w \freq}{\sqrt{2}} \t*{\mathscr K}{_m^-}
		\end{pmatrix*}
		\,,
	\end{align}
\end{widetext}
where the following abbreviations were used:
\begin{align}
	\t*{\mathscr J}{_m}
		&= \frac{J_m'(u)}{u J_m(u)}\,,
	&
	\t*{\mathscr K}{_m}
		&= \frac{K_m'(w)}{w K_m(w)}\,,
	\\
	\t*{\mathscr J}{_m^\pm}
		&= \t*{\mathscr J}{_m} \mp m / u^2\,,
	&
	\t*{\mathscr K}{_m^\pm}
		&= \t*{\mathscr K}{_m} \mp m / w^2\,.
\end{align}
The parameters $u$ and $w$ arising here are defined in \cref{eq:def u w}.
The determinant of the matrix $\hOpD{m}$ factorizes as
\begin{align}
	\det \hOpD{m}
		&= \sqrt 2 u^2 w^2
			\t*{\mathcal D}{_{\text{g,}}_m^2}
			\t*{\mathcal D}{_{\text{ph,}}_m}
\end{align}
Here, the factor whose vanishing indicates gauge and ghost modes takes the form
\begin{align}
	\t*{\mathcal D}{_{\text{g,}}_m}
		&= u^2 \t*{\mathscr J}{_m} - w^2 \t*{\mathscr K}{_m}\,,
\end{align}
and the factor whose roots correspond to physical modes can be written as
\begin{align}
	\label{eq:dispersion relation Dph}
	\t*{\mathcal D}{_{\text{ph,}}_m}
		&= (\t*{\mathscr J}{_m} + \t*{\mathscr K}{_m})(n_1^2 \t*{\mathscr J}{_m} + n_2^2 \t*{\mathscr K}{_m})
		 - \bar m^2\,,
\end{align}
where $\bar m = m \neff v^2 / (u^2 w^2)$.
The expressions arising here are the same as for straight optical fibers in flat space-time \cite{2022PhRvA.106f3511M}.

\section{Derivation of Transport Laws}
\label[appendix]{app:transport equations}

\Cref{eq:reference vectors} allows for arbitrary rescaling of the vectors $\t{\boldsymbol{\hat q}}{^\pm}$ but the interpretation of the multiplying amplitudes $\t*{\mathcal A}{^{(j)}_\pm}$ is significantly simplified if the vectors $\t{\boldsymbol{\hat q}}{^\pm}$ are normalized consistently.
As shown in Ref.~\cite[eq.~(69a)]{2022PhRvA.106f3511M}, these two vectors can be chosen such that one of them can be obtained from the other by a permutation of the components and some changes of sign.
The specific details of such a normalization play no significant role however: both $(\t{\boldsymbol{\hat q}}{^+})^\core_\flat = (\t{\boldsymbol{\hat q}}{^-})^\core_\sharp = 1$ (as is natural when taking the weak-guidance limit) and $(\t{\boldsymbol{\hat q}}{^\pm})^\core_0 = 1$ lead to the same numerical values in \cref{eq:transport phase perturbation,eq:transport polarization}.

Assuming such a symmetric normalization of the vectors $\t{\boldsymbol{\hat q}}{^\pm}$, the solvability conditions \eqref{eq:condition solvability at order j} for $j = 1$ and $j = 2$ can be written in the form
\begin{align}
	L\, \t*{\mathcal A}{_A^{(0)}}
		&= \t\Coupling{_0} \t*{\mathcal A}{_A^{(0)}} \,,
	\\
	L\, \t*{\mathcal A}{_A^{(1)}}
		&= \t{\Coupling}{_0} \t*{\mathcal A}{_A^{(1)}}
		+ \t{\Coupling}{_A^B} \t*{\mathcal A}{_B^{(0)}}\,,
\end{align}
where $L = \frac{\d}{\d\varsigma} \pm \i \ell_0 \t\beta{^{(1)}}$ is a differential operator involving the first-order term in the expansion \eqref{eq:ansatz beta}, and $\t\Coupling{_0}$ is a coefficient depending on the specific normalization of $\t{\boldsymbol{\hat q}}{^\pm}$ (in both normalization schemes mentioned above, one has $\t\Coupling{_0} = \t\coupling{_0}(v) \mathring\lapse'(\varsigma) / \mathring\lapse(\varsigma)$ for some function $\t\coupling{_0}$ that can be determined numerically).
Using computer algebra systems – an implementation in \emph{Wolfram Language} is available online \cite{mieling_2024_14142452} – one finds that the coupling coefficients $\t{\Coupling}{_A^B}$ can be written as
\begin{align}
	\begin{split}
		\t{\Coupling}{_A^B}
		={}&
		\t{\bar\Coupling}{_A^B}
		\mp \i (\ell_0 \t\beta{^{(2)}}) \t*\delta{_A^B}
		\\&
		+ [
			  \t\coupling{_1}\!(v) \ell_0 \t\beta{^{(1)}} \t{\mathring\lapse}{^\prime}/\mathring\lapse 
			+ \t\coupling{_2}\!(v) \ell_0 \t\beta{^{(1)}} \t{\mathring\lapse}{^\prime^\prime}/\t{\mathring\lapse}{^\prime}
		] \t*\delta{_A^B}
		\,,
	\end{split}
\end{align}
where $\t{\bar\Coupling}{_A^B}$ has the following structure.
Denoting by $\t\metric{_A_B} = \t\metric{_i_j} \t*\frameF{_A^i} \t*\frameF{_B^j}$ the components of the transverse metric in the complex frame $(\t*\frameF{_A^i}) = (\t*\frameF{_+^i}, \t*\frameF{_-^i})$, the matrix $\t{\bar\Coupling}{_A_B} = \t{\bar\Coupling}{_A^D} \t\metric{_D_B}$ can be written as
\begin{align}
	\t{\bar\Coupling}{_A_B}
		&= \sum_\tau \t\coupling{_A_B}(v) \t\tau{_A_B}(\varsigma)\,,
\end{align}
where $\tau$ ranges over all terms derived from the rescaled quantities $\mathring\lapse$, $\t{\mathring\lapse}{_,_\alpha}$, $\t{\mathring\lapse}{_,_\alpha_\beta}$, $\t{\mathring\shift}{_\alpha}$, $\t{\mathring\shift}{_\alpha_,_\beta}$, $\t{\mathring\pnormal}{_\alpha}$, $\t{\mathring\riemann}{_\alpha_\beta_\gamma_\delta}$ and $\t{\mathring\riemann}{_\alpha_\sigma_\beta_\sigma}$ (as defined at the end of \cref{sec:setup}) that have the appropriate Fourier index. For example, the sum for $\t\Coupling{_+_+}$ involves, i.a., the terms $(\t{\mathring\pnormal}{_+})^2$, $(\t{\mathring\lapse}{_,_+}/{\mathring\lapse})^2$, $\t{\mathring\lapse}{_,_+_+}/{\mathring\lapse}$, and $\t{\mathring\riemann}{_+_\sigma_+_\sigma}$.
The values of the coefficients $\t\coupling{_A_B}$ must generally be computed using numerical integration.

Factorizing the amplitudes as $\t*{\mathcal A}{_A^{(j)}} = \amplitude \t*\jones{_A^{(j)}}$ where the overall amplitude satisfies $\d\amplitude / \d\varsigma = \t\Coupling{_0} \amplitude$, one obtains
\begin{align}
	\label{eq:transport jones raw}
	L\, \t*\jones{_A^{(0)}} &= 0\,,
	&
	L\, \t*\jones{_A^{(1)}} &= \t\Coupling{_A^B} \t*\jones{_B^{(0)}}\,.
\end{align}
The amplitude $\amplitude$ exhibits a certain “gauge redundancy” that arises from the fact that the vectors $\boldsymbol{q}^\pm$ can be rescaled by any non-zero function $\lambda$ as $\boldsymbol q^\pm \to \lambda \boldsymbol q^\pm$. Such a transformation is equivalent to a rescaling of the multiplying amplitude by $\amplitude \to \lambda \amplitude$, which induces the transformation $\t\Coupling{_0} \to \t\Coupling{_0} + \lambda^{-1} \d \lambda/\d\varsigma$.
Since \cref{eq:transport jones raw} is unaffected by this ambiguity, the following analysis focuses on the quantities $\t*\jones{_A^{(0)}}$ and $\t*\jones{_A^{(1)}}$.

The transport equations \eqref{eq:transport jones raw} can be combined by defining $\t*\jones{_A} = \t*\jones{_A^{(0)}} + \varepsilon \t*\jones{_A^{(1)}} + O(\varepsilon^2)$, as can be motivated using renormalization methods described in Refs.~\cite{1994PhRvL..73.1311C,1996PhRvE..54..376C}, to yield
\begin{align}
	L\, \t\jones{_A}
		&= \varepsilon \t\Coupling{_A^B} \t*\jones{_B}
		+ O(\varepsilon^2)\,.
\end{align}
As shown in Ref.~\cite{2023PhRvR...5b3140M}, the two quantities $\t*\jones{_+}$ and $\t*\jones{_-}$ can be identified with projections of a complex polarization vector $\t\jones{_i}$ onto the frame $\t*\frameF{_\pm}$ via $\t\jones{_A} = \t\jones{_i} \t*\frameF{_A^i}$.
Now, the parameters $\t\beta{^{(1)}}$ and $\t\beta{^{(2)}}$ can be chosen arbitrarily without altering the physical content of the equations [this is because the splitting of the field in \cref{eq:ansatz amplitude phase} into amplitudes and a phase factor is not unique].
Particularly simple equations are obtained by setting $\t\beta{^{(1)}} = 0$ and choosing $\t\beta{^{(2)}}$ such that $\t\Coupling{_A^B}$ becomes trace-free.
This leads to the result
\begin{align}
	\label{eq:app:beta 2}
	\begin{split}
		\ell_0 \t\beta{^{(2)}}
			={}&
			  \t{\tilde\gamma}{_1} \t{\mathring\riemann}{_+_-_+_-}
			+ \t{\tilde\gamma}{_2} \t{\mathring\riemann}{_\varsigma_\varsigma}
			+ \t{\tilde\gamma}{_3} \t{\mathring\lapse}{_,_+_-} / \lapse
			\\&
			+ \t{\tilde\gamma}{_4} \t{\mathring\lapse}{_,_\varsigma}/\lapse
			+ \t{\tilde\gamma}{_5} \t{\mathring\lapse}{_,_\varsigma_\varsigma}\lapse/\lapse
			+ \t{\tilde\gamma}{_6} \t{\mathring\pnormal}{_+} \t{\mathring\pnormal}{_,_-}
			\\&
			+ \t{\tilde\gamma}{_7} \t{\mathring\lapse}{_,_+} \t{\mathring\lapse}{_,_-} / {\mathring\lapse}^2
			+ \t{\tilde\gamma}{_8} \t{\mathring\pnormal}{_(_+} \t{\mathring\lapse}{_,_-_)} / \mathring\lapse
			\,,
	\end{split}
	\\
	\label{eq:app:C(AB)}
	\begin{split}
		\t\Coupling{_A_B}
			={}&
			\varpi (\t{\mathring\shift}{_A_,_B} - \t{\mathring\shift}{_B_,_A})
			\\&
			+ \t{\tilde\eta}{_1}(\t{\mathring\pnormal}{_A} \t{\mathring\pnormal}{_B} - \t\metric{_A_B} \t{\mathring\pnormal}{_+} \t{\mathring\pnormal}{_-})
			\\&
			+ \t{\tilde\eta}{_2}(\t{\mathring\lapse}{_,_A} \t{\mathring\lapse}{_,_B} - \t\metric{_A_B} \t{\mathring\lapse}{_,_+} \t{\mathring\lapse}{_,_-}) / {\mathring\lapse}^2
			\\&
			+ \t{\tilde\eta}{_3}(\t{\mathring\pnormal}{_(_A} \t{\mathring\lapse}{_,_B_)} - \t\metric{_A_B} \t{\mathring\pnormal}{_(_+} \t{\mathring\lapse}{_,_-_)})/{\mathring\lapse}
			\\&
			+ \t{\tilde\eta}{_4} (\t{\mathring\riemann}{_A_\varsigma_B_\varsigma} - \half \t\metric{_A_B} \t{\mathring\riemann}{_\varsigma_\varsigma})
			\\&
			+ \t{\tilde\eta}{_5} (\t{\mathring\lapse}{_,_A_B} - \t\metric{_A_B} \t{\mathring\lapse}{_,_+_-}) / {\mathring\lapse}
			\,.
	\end{split}
\end{align}
The dimensionless coefficients arising here [which generally depend on the normalized frequency $v$ and the sign in \cref{eq:ansatz phase}] can be determined by numerically implementing the scheme described in \cref{sec:perturbative scheme}, in which the homogeneous solutions given in \cref{eq:solution:homogeneous} are evaluated on-shell by numerical root-finding of \cref{eq:dispersion relation Dph}, and inserted into the source terms $\t*\Sigma{^{(j,m)}_\bulk}$ and $\t*\Sigma{^{(j,m)}_\intf}$ entering \cref{eq:recursive system iteration} that can be computed symbolically using computer algebra systems (the explicit expressions are extensive and not particularly illuminating).

Numerically, one finds that $\t{{\tilde \gamma}}{_6} \approx \t{{\tilde \gamma}}{_7} \approx \half \t{{\tilde \gamma}}{_8}$ (to within \SI{0.5}{\percent} relative error) and $\t{{\tilde\eta}}{_1} \approx \t{{\tilde\eta}}{_2} \approx \half \t{{\tilde\eta}}{_3}$ (to within \SI{1}{\percent} relative error) so that the contributions of the rescaled principal normal $\t{{\mathring\pnormal}}{_A}$ and the rescaled logarithmic lapse derivative $\t{{\mathring \lapse}}{_,_A} / \mathring \lapse$ can be expressed (to that level of accuracy) via a coupling to $\t{{\mathring z}}{_A} = \t{{\mathring\pnormal}}{_A} + \t{{\mathring \lapse}}{_,_A} / \mathring \lapse$.
Similarly, one finds $\t{{\tilde\eta}}{_4} \approx \t{{\tilde\eta}}{_5}$ (to within \num{3e-5} absolute error; relative errors are undefined due to zeros of these coefficients) so that the terms involving the spatial Riemann tensor $\t{\riemann}{_i_j_k_l}$ and second logarithmic lapse derivatives $\t{\lapse}{_,_i_j} / \lapse$ in \cref{eq:app:C(AB)} can be expressed in terms spatial components of the space-time Weyl tensor $\t\Weyl{_\mu_\nu_\rho_\sigma}$ since
\begin{align}
	\begin{split}
		\t{\mathring\Weyl}{_A_\varsigma_B_\varsigma} - \t{\mathring\metric}{_A_B} \t{\mathring\Weyl}{_+_\varsigma_-_\varsigma}
		={}&
		\half (\t{\mathring\riemann}{_A_\varsigma_B_\varsigma} - \t{\mathring\metric}{_A_B} \t{\mathring\riemann}{_+_\varsigma_-_\varsigma})
		\\&
		+ \half (\t{\mathring\lapse}{_,_A_B} - \t\metric{_A_B} \t{\mathring\lapse}{_,_+_-}) / {\mathring\lapse}\,.
	\end{split}
\end{align}
Finally, the numerical analysis shows that $\t{\tilde \gamma}{_1} \approx - \half \t{\tilde \gamma}{_2}$ to within \SI{0.4}{\percent} relative error.

The transport equations \eqref{eq:transport phase perturbation} and \eqref{eq:transport polarization} are then obtained by transforming the rescaled qualities $\mathring \varphi$ entering \cref{eq:app:beta 2,eq:app:C(AB)} to their original form (as described at the end of \cref{sec:setup}) and using the fact that $\frac{\d}{\d s} = (\varepsilon / \ell_0) \frac{\d}{\d \varsigma}$.

Some of the “numerical coincidences” described above can be explained by the conformal invariance of Maxwell’s equations.
By rescaling the space-time metric by $(1 - \t x{^\alpha} \t\lambda{_\alpha}(s))^2$, where $\t\lambda{_\alpha}(s)$ are arbitrary functions, first-order terms in the expansion of the spatial metric component $\t\metric{_s_s}$ in \cref{eq:spatial metric expansion} and the lapse $\lapse$ in \cref{eq:lapse expansion} transform as $\t\pnormal{_\alpha} \to \t\pnormal{_\alpha} + \t\lambda{_\alpha}$ and $\t\lapse{_,_\alpha} \to \t\lapse{_,_\alpha} - \t\lambda{_\alpha} \lapse$.
Choosing $\t\lambda{_\alpha} = \t\lapse{_,_\alpha} / \lapse$ thus eliminates first transverse derivatives of the lapse while simultaneously replacing $\t \nu{_\alpha} \to \t z{_\alpha} \equiv \t\nu{_\alpha} + \t\lapse{_,_\alpha} / \lapse$ in the first-order expansion of the $\t\metric{_s_s}$.
Under such a conformal transformation, all second-order terms in \cref{eq:spatial metric expansion,eq:lapse expansion} are also transformed, but as shown in \cref{fig:coupling moments}, their effects are multiple orders of magnitude smaller than those arising from the linear terms under consideration.
The last numerical coincidence listed above, however, does not seem to arise from conformal invariance as the sectional curvature $K = \t\riemann{_x_y_x_y} = - \t\riemann{_+_-_+_-}$ and second lapse derivatives do not directly combine to yield a component of the Weyl tensor.

\bibliography{bibliography,bibliography_cyrillic}
\end{document}